\documentclass[aps,pre,reprint,groupedaddress]{revtex4-2}
\usepackage[utf8]{inputenc}
\usepackage[colorlinks,linkcolor=blue,citecolor=magenta,urlcolor=blue]{hyperref}
\usepackage{amsmath}
\usepackage{diagbox} 
\usepackage{graphicx}
\usepackage{amssymb}
\usepackage{mathrsfs}
\usepackage{xcolor}
\usepackage{bbold}
\usepackage{wrapfig}
\usepackage{framed}
\usepackage{mathtools}
\usepackage{sidecap}
\usepackage{xargs}
\usepackage{makecell}

\newcommand{\deltace}{\delta}
\newcommand{\deltacei}{\delta_{\mathcal{I}}}
\newcommand{\deltaceii}{\delta_{\mathfrak{D}}}

\newcommand{\rhop}{\rho_\mathrm{p}}
\newcommand{\rhoh}{\rho_\mathrm{h}}
\newcommand{\rhot}{\rho_\mathrm{t}}

\newcommand{\st}[1]{#1^{\mathrm{st}}}
\newcommand{\thm}[1]{#1^{\mathrm{th}}}

\newcommand{\drst}[1]{#1^{\mathrm{dr,st}}}
\newcommand{\Drst}[1]{#1^{\mathrm{Dr,st}}}

\newcommand{\C}{\mathcal{C}}

\newcommand{\Cst}{\mathcal{C}^{\mathrm{St}}}

\newcommand{\tauib}{\tau_{\mathcal{I}}}
\newcommand{\taurta}{\tau_{\mathrm{RTA}}}

\newcommand{\bigO}{\mathcal{O}}
\newcommand{\dd}{{\rm d}}

\newcommand{\hst}[2]{\left(#1|#2\right)_{\mathrm{st}}}

\newcommand{\ulb}[1]{h_{#1}}

\begin{document}

\title{
Burgers equation from non-thermal stationary states in nearly-integrable gases}
\author{Paweł Lisiak}
\author{Maciej Łebek}
\email[]{maciej.lebek@fuw.edu.pl}
\author{Miłosz Panfil}
\affiliation{Faculty of Physics, University of Warsaw, Pasteura 5, 02-093 Warsaw, Poland}

\date{\today}

\begin{abstract}
When a gas of particles interacts with much a larger reservoir the density dynamics on large scales is typically governed by diffusion. We study this paradigmatic problem for weakly coupled integrable systems and show that this picture gets altered, when transport is investigated on top of long-lived non-thermal states. Remarkably, for states non-invariant under parity we find Burgers equation arising in the hydrodynamic limit. 
We explicitly compute the diffusion constant and nonlinear advective coefficient of the Burgers equation using a variant of the Chapman-Enskog theory. We find an excellent agreement between our theory and numerical simulations of a simplified model of stochastic two-body collisions.
Our conclusions are based only on Galilean invariance, existence of a small system-bath coupling parameter and a small momentum exchange between the system and the bath particles during two-body scattering.
\end{abstract}

\maketitle

\section{Introduction}
The non-equilibrium dynamics of low-dimensional quantum many-body systems attracts a lot of attention due to recent advances in experimental, numerical and analytical understanding of fundamental phenomena such as thermalisation and transport~\cite{RevModPhys.83.863,review_DAlessio_2016,Bertini2021,Calabrese2016,2016JSMTE..06.4002E,QA_Caux,Ilievski_2016,Vasseur_2016,10.21468/SciPostPhysLectNotes.20,Bertini2021, Bastianello2020, Abanin2019,Tang2018,Malvania2021,Mller2021,Cataldini2022,Jepsen2020,Scheie2021,Wei2022,Le2023,Kinoshita2006}. In one dimension, one often encounters various instances of anomalous, or non-conventional hydrodynamics. This happens for instance in momentum-conserving fluids or anharmonic chains~\cite{Forster1977,Narayan2002, vanBeijeren2012, Spohn2014}, Luttinger liquids~\cite{Giamarchi2003,Bulchandani2020} and certain integrable spin chains~\cite{Ljubotina2017,Gopalakrishnan2019s,DeNardis2019s,Bulchandani2021r}. Non-conventional transport occurs more broadly in integrable systems, due to the presence of infinite number of conservation laws. In the recent years, a special role in the field of 1D transport is indeed played by the family of such models~\cite{Takahashi1999,Korepin1993}, for which there has been recently a lot of progress in understanding their equilibration and emergent hydrodynamics, dubbed generalized hydrodynamics (GHD)~\cite{doyon_ghd,fagotti_ghd,Doyon2025}. The developed techniques cover also situations in which the integrability is weakly broken~\cite{Durnin2021a,PhysRevB.101.180302,Panfil2023,Lopez-Piqueres2021,Lopez-Piqueres2022,Lopez-Piqueres2023,Caux2019,Bastianello2020a,Bouchoule2020,Bertini2015,Bertini2016,Bastianello2021,biagetti2024generalised}. 

In such case the resulting description is similar to a kinetic theory of dilute gases~\cite{Resibois,Fouvry2020}. Namely, the state of the system is described by a local distribution of quasiparticles which propagate with state-dependent effective velocities and occasionally scatter due to integrability breaking effects. Additionally, because of the presence of interactions in integrable theory there is a quasiparticle diffusion effect~\cite{hydro_diff_prl,DeNardis2019,Gopalakrishnan2018,Hbner2025,Durnin2021}. The scattering between the quasiparticles opens a possibility for a thermalisation channel which otherwise is absent. Another thermalisation mechanism is through introducing an external potential~\cite{Cao2018,Bastianello2020b,Bagchi2023,Biagetti2024}, but to simplify the problem we do not consider it here. 
\begin{figure}[t]
    \includegraphics[scale=0.49]{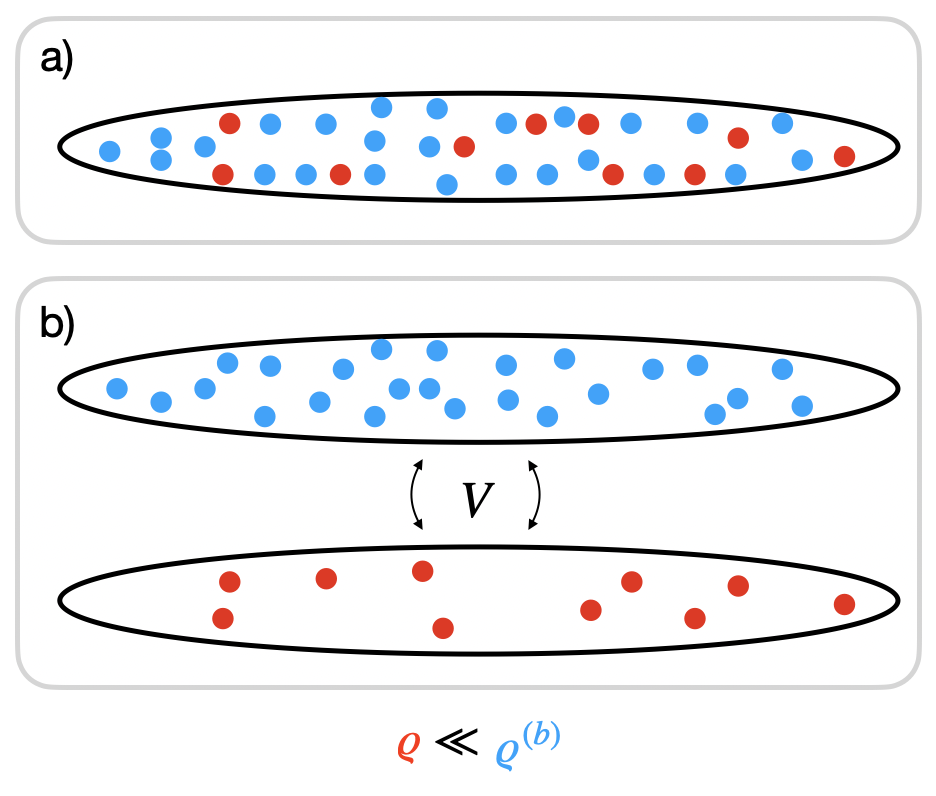}
    \caption{We consider dynamics of particles with density $\varrho$ coupled to a much larger bath with density $\varrho^{(b)} \gg \varrho$. Both systems are described by integrable models and the long-range coupling breaks the integrability. One can think about two realizations of this scenario: mixed systems (a) and long-range interacting tubes (b). Particles of the observed system feature single conservation law and we consider the problem of hydrodynamics, which emerges on large space-time scales.}
    \label{fig:intro}
\end{figure}
The observation that integrability breaking modifies the dynamics of quasiparticles similarly as weak interactions alter the dynamics of free particles can be exploited by bringing tools from the kinetic theory. In particular, such approach was used to study hydrodynamics of nearly-integrable quantum systems. An example of this idea is application of the Chapman-Enskog (ChE) theory~\cite{Resibois} to understand the emergence of conventional hydrodynamics along with computation of transport coefficients in such systems~\cite{chapmanenskog,navierstokes}. The latter was important to establish an existence of parametrically large space-time regime~\cite{Lepri2020,Chen2014,Zhao2018,Miron2019,Kundu2023,chapmanenskog}, where the conventional hydrodynamics works before giving place for anomalous one~\cite{Forster1977,Narayan2002, vanBeijeren2012, Spohn2014}. This kinetic picture can be also lifted from the quasiparticles distribution to their many-body correlations leading to a generalization of the famous Born-Bogoliubov-Green-Kirkwood-Yvon (BBGKY) hierarchy~\cite{biagetti2024generalised}. The generalized BBGKY (gBBGKY) provides a new theoretical framework to understand the integrability-breaking-induced thermalization and was proven useful in description of long-range interacting hard rod gas numerics as well as experiments in 1D dipolar gases~\cite{PhysRevX.8.021030}. 

In this work we continue following that program and consider a canonical problem of diffusion of Galilean particles. 
The mechanism for diffusion is provided by a long-range coupling to a much larger bath, also assumed to be described by a one-dimensional integrable model. As shown in Fig.~\ref{fig:intro}, one can naturally think about such scenario in the geometry of mixed systems, cf. Fig.~\ref{fig:intro}(a) or in coupled 1D tubes, cf. Fig~\ref{fig:intro}(b). The particles of the smaller system, due to long-range interactions, exchange momentum and energy with the bath and therefore the only conserved quantity associated with them is their total number. On a phenomenological level, when the bath state is assumed to be thermal, the continuity equation for the particle density $\varrho(x,t)$, combined with the gradient expansion of the current, lead to the diffusion equation
\begin{equation}\label{eq:diff_eq}
    \partial_t \varrho = \partial_x\left(\mathcal{D}(\varrho) \partial_x \varrho \right) ,
\end{equation}
with the density-dependent diffusion coefficient $\mathcal{D}(\varrho)$. The standard form of the diffusion equation is obtained by linearising the theory with the diffusion coefficient attaining a constant value. In such case the diffusion coefficient can be also computed through the Green-Kubo formula from the velocity autocorrelation function. 

The diffusion equation can be also obtained from microscopic considerations based on the kinetic theory and Boltzmann collision operator $\mathcal{I}[\rhop, \rhop^{(b)}]$ for system-bath interactions. Here $\rhop(x,\lambda,t)$ is one-body density function ($\lambda$ labels quasimomentum) and $\rhop^{(b)}$ denotes (thermal) state of the bath. 
The density is $\varrho(x,t) = \int {\rm d}\lambda\, \rhop(x, \lambda, t)$ and under the assumption $\varrho \ll \varrho^{(b)}$ the dynamics of the bath is effectively frozen out.
The diffusion coefficient can be computed with the ChE method, discussed later in the text, in Sec.~\ref{sec:nld}. It is given by an integral equation formulated in terms of the collision operator $\mathcal{I}$.

Our results generalize in various way the diffusion equation~\eqref{eq:diff_eq}. Firstly, the resulting diffusion coefficient incorporates in non-perturbative way the interactions of integrable theory, which extends the applicability of the standard kinetic theory methods to strongly correlated systems. Secondly, isolated integrable systems do not equilibrate to thermal states due to presence of an extensive number of conservation laws. While breaking the integrability through system-bath interactions ultimately thermalizes the system, if it is sufficiently weak it creates a time span in which the system might be in a prethermal quasi-stationary state~\cite{Lebek2024,biagetti2024generalised,Tang2018,Bertini2015,Bertini2016}. This leads to a new physical situation in which the particles diffuse in a non-thermal environment. Finally, if the non-thermal state is furthermore not parity invariant there are new terms appearing in the diffusion equation. These together ultimately lead to the celebrated Burgers equation (studied recently in many contexts~\cite{Kulkarni2012,Abanov2018,kim2025}) in the form
\begin{equation} \label{Burgers_intro}
   \partial_t \varrho
        + u(\varrho) \partial_x \varrho
        = \partial_x\left(\mathcal{D}(\varrho)
       \partial_x \varrho\right),
\end{equation}
where $u(\varrho)$ is advective coefficient, which can be constant or not, depending on particles' statistics. For classical statistics, we find that $u(\varrho)$ is constant, and equivalent to a simple Galilean boost of the diffusion equation~\eqref{eq:diff_eq}. The more interesting situation happens for quantum statistics, where $u(\varrho)$ is non-constant and we recover a genuine Burgers equation. This is summarized in Tab.~\ref{tab:LRN_vs_LRE}.
\begin{table}[t]
    \renewcommand{\arraystretch}{2}
	\begin{ruledtabular}
		\begin{tabular}{c c c} 
			\diagbox{state}{statistics}	& classical   &quantum    \\
            \hline
            parity-symmetric & diff. eq. \eqref{eq:diff_eq} & diff. eq. \eqref{eq:diff_eq}  \\
            \hline
            parity-asymmetric & diff. eq. \eqref{eq:diff_eq} & Burgers eq. \eqref{Burgers_intro}\\
		\end{tabular}
		\caption{Hydrodynamic equations for the system for different particle statistics and properties of the stationary state.}
        \label{tab:LRN_vs_LRE}
    \end{ruledtabular}
\end{table}
Equation~\eqref{Burgers_intro} for an almost homogeneous density profile, $\varrho \approx \varrho_0 + \delta\varrho$, after changing the reference frame to the one moving with $u(\varrho_0)$, simplifies to
\begin{equation} \label{true_Burgers_intro}
    \partial_t \delta\varrho
        + \delta \varrho \partial_x \delta \varrho
        =  \nu \partial_x^2 \delta\varrho.
\end{equation}
with time rescaled by $u'(\varrho_0)$ and $\nu = \mathcal{D}(\varrho_0)/u'(\varrho_0)$. In the literature, it is equation~\eqref{true_Burgers_intro} that is usually called Burgers equation. Here, for simplicity, we will refer to both equations~\eqref{Burgers_intro} and~\eqref{true_Burgers_intro} as Burgers equations.

We exemplify our approach by considering three different physical scenarios. First, we study a system of two coupled 1D Bose gases which mimics the experimental settings of cold atomic gases consisting of arrays of 1D systems with possibly tunable couplings~\cite{PhysRevX.8.021030}. It is known that the collision integral for such systems supports athermal stationary states~\cite{Lebek2024}. This provides a concrete physical realisation for testing the ideas presented in this work. Next, we abstract from a concrete realization by considering a generic collision integral in the relaxation time approximation (RTA) with respect to athermal state and with particle number being the only conserved quantity. Within the RTA the ChE equation for the diffusion constant can be solved. This allows us to draw general conclusions for diffusion processes in such circumstances and show how Burgers equations arise for athermal and parity-breaking background states. Finally, we consider a simple stochastic model of classical or fermionic particles colliding with bath particles at constant rate. This serves as a minimal model in which all the phenomena seen in a more sophisticated situations are present, but without technical complications. This allows us to pinpoint the sufficient conditions required to see the Burgers equation. At the same time, this model can be simulated efficiently and we successfully test the effective theory given by~\eqref{Burgers_intro} against the molecular dynamics. 

The paper is organized in the following way. In Sec.~\ref{sec:background} we provide the necessary ingredients of the kinetic theory of nearly-integrable system, specifically we discuss in the details the Thermodynamic Bethe Ansatz (TBA), the GHD equation supplemented with the Boltzmann term  and the Generalized Gibbs Ensemble (GGE) formulation of the quasiparticle distribution function. In Section~\ref{sec:athermal_states} we discuss the collision integrals supporting athermal states. In Sec.~\ref{sec:nld} we use the ChE method to derive from the GHD-Boltzmann equation the Burgers equation. In Sec.~\ref{sec:applications}, we extract the characteristic relaxation time from the collision integral for two coupled 1D gases which then serves as the input for the RTA. We compute the diffusion coefficient in the latter approximation and link it with thermodynamic properties of the gas also in athermal states. We also show how Burgers equation arises for states without parity invariance. In Sec.~\ref{sec:numerics} we demonstrate that our predictions match the microscopic simulations of the stochastic models. We relegated some more technical aspects of our work to the Appendices, to which we refer when needed.

\section{Preliminaries} \label{sec:background}

We consider a system described by the following Hamiltonian
\begin{equation} \label{Hamiltonian}
    H = H_{\mathrm{int}} +H_{\mathrm{int}}^{(b)} + \alpha H_{\mathrm{pert}}, \qquad \alpha \ll 1,
\end{equation}
where $H_{\mathrm{int}}$ and $H_{\mathrm{int}}^{(b)}$ are some integrable Hamiltonians modelling system and bath, respectively and $\alpha H_{\mathrm{pert}}$ is some small integrability-breaking perturbation
Hamiltonian (modelling system-bath interaction). The system is assumed to be Galilean-invariant.
We consider only the case when $H_{\mathrm{pert}}$ does not depend on position and time explicitly.
The dynamics of nearly-integrable models, after the initial decoherence phase, is effectively described by kinetic theory~\cite{Tang2018,biagetti2024generalised,Durnin2021,Panfil2023}. This kinetic theory is formulated in terms of the space-time dependent distribution function of quasiparticles $\rhop(x,\lambda, t)$ with quasimomentum (rapidity) $\lambda$. To understand the basics of quasiparticle description we review first the TBA and GHD of integrable models.
\subsection{Basics of TBA and GHD}
Integrable systems are characterized by infinite number of local conserved charges $Q_k$ fulfilling $[H_{\rm int}, Q_k]=0$.
Their expectation values for a state described by quasiparticle distribution $\rhop$ are
\begin{equation}
    \langle Q_k \rangle = \int \dd \lambda\, \ulb{k}(\lambda) \rhop(\lambda).
\end{equation}
For Galilean invariant models of particles with unit mass, as is assumed here, it is convenient to choose $\ulb{k}(\lambda) = \lambda^k/k!$ which implies that $Q_k$ are ultra-local charges.  Specifically, for total momentum $\ulb{1}(\lambda) = \lambda$ and for total energy $\ulb{2}(\lambda) = \lambda^2/2$. The quasiparticle picture is based upon the assumption of relaxation to GGE~\cite{Yang1969,Mossel2012} with the density matrix $\rho_{\rm GGE} \propto \exp(- \sum_j \beta_j Q_j)$, where $\beta_j$ are Lagrange multipliers.

We introduce now more ingredients of the TBA in simlar way as presented in~\cite{lecture_notes_ghd}. We start with free energy density $\mathsf{f}=\frac{1}{2 \pi} \int {\rm d} \lambda F[\epsilon(\lambda) ]$, where $\epsilon(\lambda)$ is so-called pseudoenergy and function $F$ depends on particles' statistics (for fermionic particles  $F[\epsilon(\lambda) ] = - \log(1+e^{-\epsilon(\lambda)})$). The pseudoenergy is determined from the following equation
\begin{equation} \label{eq:gge_condition}
    \epsilon(\lambda) = \epsilon_0(\lambda) + \int {\rm d} \lambda' \mathcal{T}(\lambda-\lambda') F[\epsilon(\lambda')],
\end{equation}
with bare pseudoenergy given by $\epsilon_0(\lambda) = \sum_j \beta_j h_j(\lambda)$. Moreover, by $\mathcal{T}(\lambda-\lambda')$ we denoted model-dependent scattering kernel. The set $\{\beta_j\}_{j=0,1,\ldots}$ specifies the state. From pseudoenergy we can compute statistical factor $f(\lambda) = ({\rm d}^2F/{\rm d}\epsilon^2)/({\rm d}F/{\rm d}\epsilon)$ and occupation function $n={\rm d}F/{\rm d} \epsilon$ defined also as $n=\rhop/\rho_{\rm t}$, where $\rho_{\rm t}$ is the total density of states. It fulfills the following integral equation
\begin{equation}
    \rho_{\rm t}(\lambda) = \frac{1}{2\pi} + \int {\rm d} \lambda' \mathcal{T}(\lambda-\lambda') \rhop(\lambda').
\end{equation}
Next, we introduce dressing operations. The small dressing is defined through integral equation
\begin{equation}
    g^{\rm dr}=g+\mathcal{T}n \cdot g^{\rm dr}.
\end{equation}
The big Dressing is the result of the action of the operator $\mathbb{F}$, $g^{\rm Dr} = \mathbb{F}^t g$ ($t$ denotes transposition) with
\begin{equation} \label{Dr_oper}
    \mathbb{F}(\lambda,\mu) = \delta(\lambda -\mu) + \partial_{\lambda} \left(n(\lambda) F(\lambda|\mu)\right).
\end{equation}
Above we introduced backflow function $F(\lambda|\mu)$~\cite{Korepin1993}, appearing naturally in studies of excitations in integrable systems. The two dressings are related
\begin{equation} \label{dressing_relation}
    \left(g^{\rm Dr}\right)' = (g')^{\rm dr}.
\end{equation}
For future reference we introduce Dressed momentum $k(\lambda)=h_1^{\rm Dr}(\lambda)$ and energy $\omega(\lambda)=h_2^{\rm Dr}(\lambda)$. We also mention useful relation that $k'(\lambda)=2 \pi \rho_{\rm t}(\lambda)$.

We move on and discuss ingredients of GHD~\cite{lecture_notes_ghd}. The effective velocity $v$ of quasiparticles obeys the integral equation
\begin{equation} \label{eq:eff_v_int_eq}
    v(\lambda) = \ulb{1}(\lambda) + 2 \pi \int \dd \lambda' \mathcal{T}(\lambda - \lambda')
        \rhop(\lambda') \left(v(\lambda') - v(\lambda)\right).
\end{equation}
Alternatively, the effective velocity can be written as $v(\lambda)=h_1^{\rm dr}(\lambda)/h_0^{\rm dr}(\lambda)$. For later use it will be convenient to introduce kernels of hydrodynamic matrices: flux Jacobian
\begin{equation} \label{flux_Jacobian}
    \mathcal{A} =(\mathbf{1}-n\mathcal{T})^{-1} v (\mathbf{1}-n\mathcal{T}),
\end{equation}
along with charge and current susceptibility matrix kernels
\begin{align}
    \mathcal{C} &= \left(\mathbf{1} - n \mathcal{T}\right)^{-1} \rhop f
        \left(\mathbf{1} - \mathcal{T} n\right)^{-1}, \label{eq:C_matrix} \\
    \mathcal{B} &= \left(\mathbf{1} - n \mathcal{T}\right)^{-1} v\,
        \rhop f \left(\mathbf{1} - \mathcal{T} n\right)^{-1} \label{eq:B_matrix}.
\end{align}
The last ingredient is the diffusion operator, which we denote by $\mathfrak{D}_\rho = \mathfrak{D}_\rho(\lambda, \mu; x, t)$. It is state-dependent and expressible through objects already introduced. For specific form of $\mathfrak{D}$ and discussion of its properties we refer to Appendix~\ref{app:diffusion}.
\subsection{GHD-Boltzmann equation}
We will assume here that the dynamics of integrable systems in the presence of integrability-breaking effects is given by GHD-Boltzmann equation~\cite{Durnin2021a,Panfil2023}
\begin{equation} \label{eq:ghd_boltzmann}
    \partial_t \rhop
            + \partial_x(v_{\rho} \, \rhop) = \frac{1}{2} \partial_x \left(
                \mathfrak{D}_{\rho} \partial_x \rhop \right)
            + \mathcal{I}[\rhop].
\end{equation}
where $\mathcal{I}[\rhop]$ is the collision operator responsible for interaction between the observed system and the bath. We indicated the dependence on state in effective velocity and diffusion operator by putting $\rho$ in the lower index. Moreover, we assume that the bath system is much larger and in consequence, does not evolve in time. Equation \eqref{eq:ghd_boltzmann} describes the system only and thus does not conserve momentum and energy. However, it does conserve particle number, which is reflected in collision integral
\begin{equation} \label{eq:boltzmann_conservation}
    \int \dd\lambda\, \ulb{0}(\lambda) \mathcal{I}[\rhop] = 0.
\end{equation}
We also assume that $\mathcal{I}[\rhop]$ has a family of stationary states $\rhop^{\rm st}$ parametrised by their density $\varrho$. We will give explicit examples of collision integrals with such properties in the following section.

\section{Athermal stationary states} \label{sec:athermal_states}

The aim of this section is to provide a construction of collision integrals that support athermal stationary states and to expose relevant physics behind them. We can focus here on the homogeneous systems. The generalization to inhomogeneous systems follows then from general arguments of separation of scales.

Athermal states relevant to the present context were first found in the setup of coupled Lieb-Liniger (LL) gases~\cite{Lebek2024}. In this work, thanks to the recent development of gBBGKY~\cite{biagetti2024generalised}, we present arguments which hold for arbitrary weakly coupled integrable systems. In gBBGKY for coupled integrable systems we assume that the system and bath interact through long-range density-density interaction given by
\begin{equation}\label{eq:pot_GBBGKY}
    V(x) =\frac{V_0}{\xi}\varphi \left(\frac{x}{\xi} \right),
\end{equation}
where $\xi$ is the characteristic range of the potential, $V_0$ measures its strength and function $\varphi$ gives the interaction profile. Importantly, the parameter $\xi$ should be thought of as a large parameter in the theory in order to provide a mechanism for truncation of the gBBGKY hierarchy. Moreover, we will assume here that the bath does not evolve, which can be realized for instance by a large density imbalance between the two systems. The aforementioned kinetic framework predicts that the dynamics at sufficiently late times of the system is given by Boltzmann-like kinetic equation.
\begin{equation}
    \partial_t \rhop(\lambda) = \mathcal{I}_{2p}(\lambda)\left(1 + \mathcal{O}(\xi^{-1})\right),
\end{equation}
where collision integral $\mathcal{I}_{2p}$ describes two-particle scattering processes, which will be discussed in the following. There are important corrections, written above as $\mathcal{O}(\xi^{-1})$, which involve collision operators responsible for three and more particle collisions. These processes are relevant on timescales longer than $\mathcal{I}_{2p}$ and  eventually thermalize the system. In what follows, we neglect them focusing on timescales relevant for $\mathcal{I}_{2p}$. Let us also note that the same picture emerges from Fermi's Golden Rule (FGR) treatment of integrability breaking as developed in~\cite{Durnin2021a,Panfil2023}. There, the collision operator involve terms with more and more particle collision processes, with the two-particle scatterings dominating in the system-bath setup. The exact equivalence of between FGR and gBBGKY was established in~\cite{biagetti2024generalised}.

To understand the simplest relevant physical scenario and minimal ingredients for athermal states let us abstract from the specific form of $\mathcal{I}_{2p}$ arising from a microscopic theory. We start broadly and consider general Master equation for two-particle scatterings between classical system and classical bath
\begin{equation}\label{eq:Master_general}
\begin{aligned}
    \mathcal{I}[\rhop](\lambda)&= \int \dd \lambda' \dd \lambda_b \dd \lambda_b' \mathcal{W}(\lambda,\lambda_b;\lambda',\lambda_b')  \times\\
    &\delta(k) \delta(\omega)\left(\rhop(\lambda)\rhop^{(b)}(\lambda')-\rhop(\lambda')\rhop^{(b)}(\lambda)\right),
\end{aligned}
\end{equation}
where conservation laws for a scattering process are imposed by Dirac deltas with $k=k(\lambda)+k_b(\lambda_b)-k(\lambda')-k_b(\lambda'_b)$ and $\omega =\omega(\lambda)+\omega_b(\lambda_b)-\omega(\lambda')-\omega_b(\lambda'_b) $. In general, we assume that quasiparticle with quasimomentum $\lambda$ carry momentum $k(\lambda)$ and energy $\omega(\lambda)$. In our notation $k, k_b$ are momenta of system and bath particles, respectively, the same goes for energies $\omega, \omega_b$. The quantity $\mathcal{W}(\lambda,\lambda_b;\lambda',\lambda_b')$ is naturally related to a rate of scattering $\lambda' + \lambda'_b \to \lambda + \lambda_b$.

Importantly, we assume that dispersion relations in both systems are not necessarily the same (meaning $\omega_b \neq \omega$ and $k_b \neq k$), which in the context of classical particles can be realized for instance by different masses, or different length of the rods (in the case of hard rod gas). Non-equal dispersions generically prohibit~\cite{Lebek2024} the existence of athermal state and thermal states (given by Maxwell-Boltzmann distributions) are the only fixed points of \eqref{eq:Master_general}.

The complementary situation of equal dispersion happens when both gases consist of identical free particles with $k(\lambda)=\lambda$ and $\omega(\lambda)=\lambda^2/2$. The momentum-energy conservation laws imply then that scattering amounts to permutation of velocities meaning that $\lambda=\lambda_b'$ and $\lambda_b=\lambda'$. We readily find the stationary state in such case which is
\begin{equation}\label{eq:Master_stat}
    \rhop^{\rm st}(\lambda) = \frac{\varrho}{\varrho^{(b)}} \rhop^{(b)}(\lambda).
\end{equation}
Obviously, if $\rhop^{(b)}$ is athermal then the state above is athermal as well. The case of two classical gases with equal dispersions is the simplest example of dynamics leading to non-thermal stationary state.

Remarkably, the conclusion about lack of non-thermal fixed points in the case of non-equal dispersions changes in the \textit{grazing limit} of the equation \eqref{eq:Master_general}, which amounts to restricting possible scattering processes to only these, where the momentum exchange is small. In that limit, the collision integral acquires the form of Landau-like kinetic equation~\cite{Resibois,Balescu1997}

\begin{equation}\label{eq:Igrazing}
\begin{aligned}
    \mathcal{I}^{\rm gr}[\rhop](\lambda) &= -\partial_\lambda \int \dd \lambda_b \tilde{\mathcal{W}}(\lambda,\lambda_b) \delta(v(\lambda)-v_b(\lambda_b)) \times\\
    &\rhop(\lambda) \rhop^{(b)}(\lambda_b) \left[\epsilon'(\lambda)-\frac{k'(\lambda)}{k'_b(\lambda_b)}\epsilon_b'(\lambda_b)\right],
\end{aligned}
\end{equation}
with $\tilde{\mathcal{W}}(\lambda,\lambda_b)$ directly related to collision rate $\mathcal{W}$. For the precise definition of grazing limit as well as expression for $\tilde{\mathcal{W}}$ see Appendix~\ref{app:grazing}.

The stationary state of~\eqref{eq:stat_grazing} fulfils a set of two equations
\begin{equation}\label{eq:stat_grazing}
    v(\lambda)=v_b(\lambda_b), \quad \epsilon'(\lambda)= \frac{k'(\lambda)}{k'_b(\lambda_b)}\epsilon'_b(\lambda_b).
\end{equation}
Let us emphasize that this stationary state appears even though two-body scattering \textit{does not} lead to a simple momentum exchange. We also stress that the solution to the equation above is athermal in the case of athermal bath. In~\cite{Lebek2024} such athermal states were found in the setup, where both the system and bath are described by the LL models. 

In the generic interacting integrable models, the dispersion relation is a dynamic quantity, it depends on the state of the system which in turn reflects {\rm e.g.} the interactions and the density. Therefore, two such systems that are weakly coupled with potential that allows only for small exchange of momentum naturally provide a physical realisation in which two body scatterings in (1+1)D lead to momentum redistribution. In this case we expect the stationary state described by $\eqref{eq:stat_grazing}$.

Summing up, the existence of possible non-thermal states depends on dispersion relations in both systems and on grazing limit of collision operator, cf. Tab.~\ref{tab:athermal}. Lastly, let us mention, that  the found stationary states become thermal states in the case of thermal bath.

\begin{table}[t]
    \renewcommand{\arraystretch}{2}
	\begin{ruledtabular}
		\begin{tabular}{c c c} 
			\diagbox{dispersions}{collisions}	& \shortstack{at small momenta \\ (grazing limit)}   & all momenta    \\
            \hline
            same & atherm. & atherm.   \\
            \hline
            different & atherm. & therm.\\
		\end{tabular}
		\caption{Conditions for possible existence of athermal states in dynamics driven by two-particle scattering processes. Athermal states are excluded only if the dispersions of the systems $H_{\rm int}$, $H_{\rm int}^{(b)}$ are different, and the grazing limit of small momentum exchange is absent.}
        \label{tab:athermal}
	\end{ruledtabular}
\end{table}

In what follows we go beyond the case of classical particles described by \eqref{eq:Master_general} and present different treatments of weakly coupled integrable systems, quantum or classical. We start with the most concrete case of coupled LL models, which can be studied using FGR. This analysis is later generalized to arbitrary weakly coupled integrable models using the gBBGKY hierarchy. Lastly, we propose a simplified model of stochastic two-body collisions which offers particularly simple structure of collision integral, while capturing all the relevant features of the dynamics. It also has the advantage of efficient numerical simulation and analytical expressions for diffusion constants.
\subsection{Fermi's Golden Rule}
We start with the FGR treatment of nearly-integrable quantum models.
Within this approximation, the collision integral involves matrix elements of the perturbing operator. In our case this origins from the integrability breaking term in the Hamiltonian which we assume couples to the local particles' density operator $q_0(x)$. However, we can be slightly more general and assume that it couples to density $q_j(x)$ of any of the conserved charges $Q_j = \int {\rm d}x\, q_j(x)$.

The collision integral for two-particle processes takes then the following form
\begin{equation} \label{BareCollisionIntegralDef}
    \mathcal{I}[\rho_{\rm p}] = \mathbb{F} \mathcal{I}_0[\rho_{\rm p}],
\end{equation}
where the backflow operator~\eqref{Dr_oper} describes the response of the whole state on the scattering process described by the bare collision integral $\mathcal{I}_0[\rhop]$. The latter is given by the usual form of the Master equation,
\begin{equation} \label{master}
    \mathcal{I}_0[\rho_{\rm p}](\lambda) = \int {\rm d}\mu \left[w(\lambda,\mu) \rhop(\mu) \rhoh(\lambda)  - (\lambda \leftrightarrow \mu)\right],
\end{equation}
with the transition rate $w(\lambda, \mu)$ and density of holes defined as $\rho_{\rm h}=\rho_{\rm t}-\rho_{\rm p}$. The transition rate is computed from the FGR and involves matrix elements of the perturbing operator $q_j(x)$. Typically, the transition rate obeys a detailed balance relation,
\begin{equation} \label{detailed_transitions}
    w(\lambda, \mu) = e^{-\beta (\omega(\lambda) - \omega(\mu))} w(\mu, \lambda),
\end{equation}
and in such circumstances, the stationary state of~\eqref{master} is thermal with inverse temperature $\beta$. Lack of the detailed balance relation open ways for athermal stationary states. 

To illustrate a situation in which athermal stationary states exist we consider a two coupled 1D systems, see Fig.~\ref{fig:intro}(b), with total Hamiltonian
\begin{equation}\label{eq:Hamiltonian}
    H = H_1 + H_2 + V_{12}.
\end{equation}
We assume that $H_i$ are integrable, for example are given by the LL model~\cite{Lieb1963,Lieb1963a}, and $V_{12}$ breaks the integrability. In calculations, it is natural to treat system and bath on equal footing which is reflected in a slightly different notation than in Eq.~\eqref{Hamiltonian}. Experimentally relevant choice of $V_{12}$ is the long-range interaction potential between two tubes that couples to local densities
\begin{equation} \label{V_pert}
    V_{12} = \int {\rm d}x_1 {\rm d}x_2\, V(x_1 - x_2) q_i(x_1) q_j(x_2).
\end{equation}
In writing the perturbation term we allow for coupling between densities of different charges. The usual density-density interaction corresponds to $i=j=0$.

In the homogeneous system, the quasiparticles density in the two tubes evolves according to the set of equations
\begin{equation} \label{TwoTubesEvolution}
    \partial_t \rhop^{(1)} = \mathcal{I}_{12}[\rhop^{(1)}, \rhop^{(2)}], \quad
    \partial_t \rhop^{(2)} = \mathcal{I}_{21}[\rhop^{(2)}, \rhop^{(1)}],
\end{equation}
with the corresponding transition rates
\begin{align} \label{TransitionRates}
    w_{12}(\lambda, \mu) &= \tilde{V}^2(k) |F_i^{(1)}(\lambda, \mu))|^2 S_j^{(2)}(-k, -\omega), \\
    w_{21}(\lambda, \mu) &= \tilde{V}^2(k) |F_j^{(2)}(\lambda, \mu))|^2 S_i^{(1)}(-k, -\omega).
\end{align}
Here $F_i^{(1)}(\lambda, \mu)$ is the one particle-hole form factor of the operator $q_i$ evaluated in the first tube. The rapidity of the particle and the hole are $\lambda$ and $\mu$ respectively. With $k \equiv k(\lambda) - k(\mu)$ and $\omega \equiv \omega(\lambda) - \omega(\mu)$ we denote momentum and energy of the particle-hole pair. $\tilde{V}(k)$ is the Fourier transform of the interaction potential from Eq.~\eqref{V_pert}. Finally, $S_j^{(2)}(k, \omega)$ is the dynamic structure factor in the second tube, namely Fourier transform of the connected two point function $\langle q_j^{(2)}(x,t) q_j^{(2)}(0) \rangle^c$ of density $q_j$. In principle, there are also contributions to the transition rates that involve form-factors with larger number of particle-hole pairs. As shown in~\cite{Lebek2024} their contribution is subleading in the momentum transferred between the tubes and is important at larger time scales. Importantly, the higher particle-hole processes are responsible for ultimate thermalization of the system~\cite{Panfil2023}.

We consider now an imbalance limit in which density of particles in the second tube is much larger than in the first one. In the Boltzmann picture, the evolution proceeds as a succession of $2$-body scatterings. Every such scattering process has much larger impact on a distribution in a first tube and instead, the much denser second tube evolves slowly. In the self-diffusion limit we assume that it does not evolve at all and distribution $\rhop^{(2)}$ is fixed. The second tube acquires then a role of a bath. In such case the evolution in the first tube is given by equation~\eqref{master} with the transition rate
\begin{equation}
    w(\lambda, \mu) = \tilde{V}^2(k) |F_i^{(1)}(\lambda, \mu))|^2 S_j^{(2)}(-k, -\omega),
\end{equation}
If the bath is in a thermal state then its correlation function obeys the detailed balance relation
\begin{equation}
    S_j^{(2)}(k, \omega) = e^{-\beta \omega} S_j^{(2)}(-k, -\omega).
\end{equation}
In the consequence the transition rates obey~\eqref{detailed_transitions} since the form-factors are particle-hole symmetric and the stationary state is thermal. Otherwise, the stationary state of~\eqref{master} is athermal.

Note also, that for illustration purposes the derivation of the collision integral was presented in the case of two coupled integrable models with density imbalance. However, the only information that we need of the bath is its dynamic response function $S_j^{(2)}$. Instead, for the description of the system, its integrability is crucial as it provides the quasiparticle picture.

This being said, if the bath is in fact integrable, then a form of a generalized detailed balance holds~\cite{SciPostPhys.1.2.015}
\begin{equation}
    S_j^{(2)}(k, \omega) = e^{- k \mathcal{F}_2(\omega/k)} S_j^{(2)}(-k, -\omega), 
\end{equation}
where $\mathcal{F}_2(v) = \epsilon_2'(\lambda)/k_2'(\lambda)$
with $\lambda$ fixed by $v_2(\lambda) = v$. This leads then to the stationary state equations
\begin{equation} \label{stationarity}
    \frac{\epsilon_1'(\lambda)}{k_1'(\lambda)} = \mathcal{F}_2(v_1(\lambda)).
\end{equation}
Upon substituting in this equation the expression for $\mathcal{F}_2$ we recover the stationary state equation as presented in~\cite{Lebek2024} or in Eq.~\eqref{eq:stat_grazing} above. 

Following~\cite{Lebek2024} let us discuss some specific cases. First, if we specialize to a thermal state, then $\epsilon_2'(\lambda) = \beta \omega_2'(\lambda)$ and $\mathcal{F}_2(v) = \beta v$. The stationary state of the system is then thermal with the same inverse temperature $\beta$. The second special case is when states and Hamiltonians in both tubes are the same. In this cases the collision integral (at the 1 particle-hole level) is exactly zero. However, such situation cannot occur in the self-diffusion problem since the densities, and therefore states, are different by construction. Finally, in the Tonks-Girardau limit of the LL model the dressing is absent. Then the effective velocities are state independent and equal to $\ulb{1}$. The stationarity condition simplifies to $\epsilon_{1,0}'(\lambda) = \epsilon_{2,0}'(\lambda)$. Crucially, in the Tonks-Girardeau gas the form-factors with larger number of particle-hole excitations vanish and the athermal states are stationary beyond the small momentum approximation, due to equal dispersions in both systems.
\subsection{Generalized BBGKY hierarchy}

In this section we take a more general perspective on the integrability breaking and employ recently developed gBBGKY hierarchy for the dynamics of correlation functions~\cite{biagetti2024generalised}. We specialize to the system described by the Hamiltonian \eqref{eq:Hamiltonian} and assume homogenous states. Importantly, the interaction potential is assumed to be long-ranged and given by the form~\eqref{eq:pot_GBBGKY}. The full hierarchy is exposed in~\cite{biagetti2024generalised} and involves correlation functions of all orders. To understand the appearance of athermal states it is sufficient however to truncate the hierarchy at the level of two-particle (system-bath) connected correlation function $g_{\lambda_1,\lambda_2}{(x_1,x_2)}$. Such correlation is sufficient to capture two-body scattering physics discussed earlier. Obviously, there are more terms and equations in the hierarchy and they describe effects ultimately leading to thermalization. However, this will happen on a longer timescale and we neglect these effects in present discussion.  On the level of hierarchy, this means that we have to deal with three dynamical equations (for the purpose of this section, we use the notation where index $1$ refers to the system and $2$ refers to the bath and quasimomentum dependence is indicated in lower index)
\begin{equation}
\begin{split}
    \partial_t \rho^{(1)}_{\lambda_1} &= \int \dd x_2 \dd \lambda_2 V'(x_1-x_2)\partial_{\lambda_1} g_{\lambda_1,\lambda_2}(x_1,x_2),\\
    \partial_t \rho^{(2)}_{\lambda_2} &= \int \dd x_1 \dd \lambda_1 V'(x_1-x_2)\partial_{\lambda_2} g_{\lambda_1,\lambda_2}(x_1,x_2),
\end{split}
\end{equation}
\begin{equation}
\begin{aligned}
    &\partial_t g_{\lambda_1, \lambda_2}(x_1,x_2) +\\
    &\mathcal{A}_{\lambda_1,\gamma}^{(1)}\partial_{x_1}g_{\gamma,\lambda_2}(x_1,x_2)+\mathcal{A}_{\lambda_2,\gamma}^{(2) }\partial_{x_2}g_{\lambda_1,\gamma}(x_1,x_2)=\\
    &V'(x_1-x_2) \int \dd x_3 \left(\partial_{\lambda_1}\rho_{\lambda_1}^{(1)}\mathcal{C}^{(2)}_{\lambda_2,\lambda_3}-\partial_{\lambda_2}\rho_{\lambda_2}\mathcal{C}^{(1)}_{\lambda_1,\lambda_3}\right).
\end{aligned}
\end{equation}

The hierarchy can be integrated in time, as in standard derivations of collision kernels from BBGKY hierarchy~\cite{biagetti2024generalised,Fouvry2020}, leading to coupled equations of the form of~\eqref{TwoTubesEvolution}. In Appendix~\ref{app:Q-derivation} we go through this calculation. The resulting collision integrals take the Landau form in the grazing limit. The explicit form is given in \eqref{eq:BoltzmannGGBKY}. The resulting collision operator, upon specifying to LL model exactly agrees with FGR result. The athermal states featured by it are exactly of the form~\eqref{eq:stat_grazing}. The approach of gBBGKY is the most versatile as it captures both classical and quantum systems. In particular, it allowed for derivation of collision term for long-range interacting hard rods, successfully tested against exact molecular dynamics in~\cite{biagetti2024generalised}. In general, when specified to classical statistics the resulting collision integral exactly takes the form of~\eqref{eq:stat_grazing} with 
\begin{equation}
    \tilde{\mathcal{W}}(\lambda,\lambda_b) = 2 \pi^2 \int \dd k |k| \tilde{V}^2(k) (k_b'(\lambda_b))^2
\end{equation}
where $\tilde{V}(k)$ denotes Fourier transform of interaction potential~\eqref{eq:pot_GBBGKY}. For more details see Appendix~\ref{app:Q-derivation}.

\subsection{Stochastic velocity swap models}
The minimal version of two-particle scattering dynamics corresponds to putting $\mathcal{W}(\lambda,\lambda_b;\lambda',\lambda_b')= \gamma=\text{const}$ in \eqref{eq:Master_general} meaning that all possible scatterings happen at the same rate. The collision integral simplifies then to
\begin{equation}\label{eq:swap1}
    \mathcal{I}=\gamma \left(\varrho \rhop^{(b)}(\lambda)-\varrho^{(b)}\rhop(\lambda) \right).
\end{equation}
This model can be understood as an adaptation of three-particle collision model studied in~\cite{Ma1983,Miron2019} to our context. The proposed model can be efficiently simulated by imposing stochastic velocity-swapping collisions between ideal classical gases describing system and bath. Obviously, the stationary state of \eqref{eq:swap1} is given by \eqref{eq:Master_stat}.

\begin{figure}[t]
    \includegraphics[scale=0.465
]{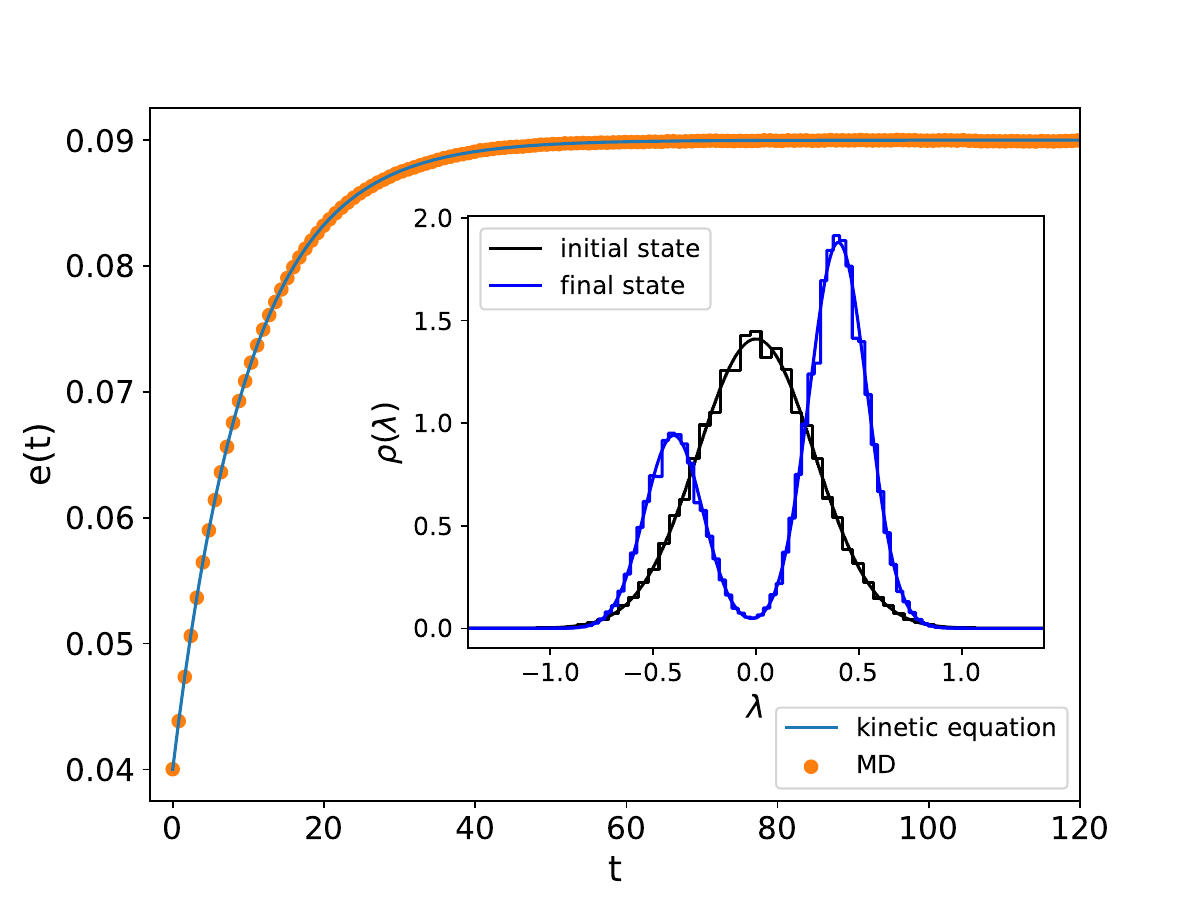}
    \caption{Relaxation dynamics of kinetic energy per particle $e(t)$ in classical stochastic velocity swap model. The molecular simulation agrees very well with solution of kinetic equation \eqref{eq:swap1}. Inset: initial and final distributions of velocities. Solid lines are analytical distributions, in particular the microscopic final state agrees very well with analytical prediction \eqref{eq:Master_stat}. We consider homogenous system of $N=1600$ particles with $\varrho=1, \varrho^{(b)}=100$ and with $\gamma=0.001$. The initial system state is  thermal with temperature $T=0.08$, whereas the bath system state is asymmetric Bragg-split state given by $\rhop^{(b)} \propto e^{-(\lambda+\lambda_B)^2/\sigma^2}+be^{-(\lambda-\lambda_B)^2/\sigma^2}$ with  $\sigma=0.2$, $\lambda_B=0.4$ and  $b=0.5$. The results were averaged over 1000 realizations.}
    \label{fig:relaxation}
\end{figure}

Due to classical statistics, the system displays standard diffusion equation~\eqref{eq:diff_eq} on large scales. As we would like to see Burgers equation~\eqref{Burgers_intro} in molecular dynamics, we extend this model to fermions interacting with classical bath, again with a constant collision rate. Master equations for scattering processes with Pauli blocking is 
\begin{equation}\label{eq:swap2}
\begin{aligned}
    &\mathcal{I}= \gamma \int {\rm d}\lambda' \times\\
    &\left[(1-n(\lambda))\rhop^{(b)}(\lambda)\rhop(\lambda')- (1-n(\lambda'))\rhop^{(b)}(\lambda')\rhop(\lambda)\right].
\end{aligned}
\end{equation}
The model features the following stationary state
\begin{equation}\label{eq:stat_ferm}
    \rhop^{\rm st}(\lambda) = \frac{1}{2 \pi } \frac{\rhop^{(b)}(\lambda)}{\rhop^{(b)}(\lambda)+A},
\end{equation}
where constant $A$ is fixed by normalization $\int \dd \lambda \rhop^{\rm st}(\lambda)=\varrho$.
The algorithm which implements classical model as well as Pauli principle in scatterings is presented in Appendix~\ref{app:algorithm}.
The models \eqref{eq:swap1} and \eqref{eq:swap2} are introduced as benchmarks for our effective, hydrodynamic theory, which can be efficiently simulated. In Fig.~\ref{fig:relaxation} we present numerical simulations of classical version of the model in homogenous setting, confirming the existence of athermal stationary states as well as the correctness of kinetic equation~\eqref{eq:swap1}. 

\section{Hydrodynamic equation} \label{sec:nld}

The effective hydrodynamic equation is obtained using the ChE method known from the kinetic theory~\cite{Resibois}. Before going into the details of the method, let us start with summarizing the most important results of this section.

The equation for density $\varrho(x,t)$ which we find on large scales in the most general case takes the form~\eqref{Burgers_intro}.
Expressions for the advective coefficient $u(\varrho)$, related to the local center-of-mass velocity of the system, and the diffusion coefficient $\mathcal{D}(\varrho)$ are
\begin{align}
\begin{split}
    u(\varrho) = \int \dd \lambda\,\ulb{1} \frac{\partial \st{\rhop}}{\partial \varrho}, \quad
    \mathcal{D}(\varrho) = -\int {\rm d}\lambda\, \ulb{1} a(\lambda, \varrho).
\end{split}
\end{align}

Here $a(\lambda, \varrho)$ is a solution to the ChE equation detailed below. The parameter $u(\varrho)$ vanishes for parity-symmetric stationary states $\rhop^{\rm st}$, leading to diffusion equation~\eqref{eq:diff_eq}. For parity-asymmetric state $\rhop^{\rm st}$, which cannot be brought to symmetric states by Galilean boost, there are two cases. For classical statistics of system particles, $u(\varrho)=\text{const}$ and it can be removed through boost. For quantum statistics, $u(\varrho)$ is density-dependent, leading to Burgers equation.

We move now to the ChE method. In papers~\cite{navierstokes, chapmanenskog} similar computation is performed for nearly-integrable system with total number of particles, momentum and energy conserved.
At the foundations of the ChE method lies the separation of scales between the microscopic collision integral relaxation time $\tau_\mathcal{I}$ and macroscopic relaxation due to $\partial_x \left(v_{\rhop} \, \rhop\right), \partial_x(\mathfrak{D}_\rho \partial_x \rhop)$ terms in the GHD equation.
The separation of scales gives us a suitable small parameter in which various quantities can be expanded.
To accommodate the GHD diffusion in the (perturbative) ChE method we must additionally
    assume separation between characteristic spatial scale of the GHD diffusion and the characteristic
    spatial scale of $\rhop$~\cite{navierstokes}.
More precisely, we demand
\begin{align}   \label{eq:ce_params_def}
\begin{split}
    \deltacei := (\tauib \langle v \rangle)/\ell_{\mathrm{h}} \ll 1, \quad 
    \deltaceii := \ell_{\mathrm{mic}}/\ell_{\mathrm{h}} \ll 1,
\end{split}
\end{align}
where $\ell_{\rm h}$ is hydrodynamic length characterizing spatial variations of $\rhop$ and
    $\langle v\rangle$ is the characteristic velocity of the state $\rhop$.
By $\ell_{\mathrm{mic}}$ we denoted characteristic spatial scale of the GHD diffusion process.
The introduced small parameters $\deltacei,\deltaceii$ are called ChE parameters.
In classical kinetic theory $\deltacei$ is known as Knudsen number.
For more discussion on the second ChE parameter we refer to \cite{navierstokes}. Another important aspect of the ChE method is the assumption, that time enters $\rhop$ only implicitly through the time-dependence of conserved charge, i.e.,
\begin{equation} \label{eq:normal_solution}
    \rhop(\lambda, x, t) = \rhop\left(\lambda, x, \,\varrho\left(x, t\right)\right).
\end{equation}
Solution to GHD-Boltzmann equation satisfying this condition is called a normal solution.

\subsection{Evaluation of ChE procedure}

We start by integrating the GHD-Boltzmann equation over $\dd \lambda$. As a result, we find the continuity equation for density
\begin{equation} \label{eq:ghd_conservation}
    \partial_t \varrho
    + \partial_x j_\rho = 0,
\end{equation}
with the following particle current
\begin{equation} \label{eq:rhop_velocity}
    j_\rho
        =  \int \dd \lambda \, v_{\rho} \, \rhop
        =  \int \dd\lambda \, \ulb{1} \, \rhop,
\end{equation}
where in the second equality we used that, for a Galilean invariant system, $v_{\rho}$ satisfies integral equation~\eqref{eq:eff_v_int_eq}, supplemented with a fact that $\mathcal{T}(\lambda)$ is symmetric~\cite{chapmanenskog}. In obtaining \eqref{eq:ghd_conservation} we also used the Markovian property of diffusion matrix, discussed in Appendix~\ref{app:diffusion}. Obviously, at this point~\eqref{eq:ghd_conservation} is not closed as the current depends on the whole distribution $\rhop$. To make some progress we have to take advantage of smallness of ChE parameters.

To see how ChE parameters~\eqref{eq:ce_params_def} appear in
the GHD-Boltzmann equation~\eqref{eq:ghd_boltzmann} we introduce dimensionless variables~\cite{navierstokes, chapmanenskog, Resibois} $\bar{t} = t/\tau_\mathrm{h}$ and $\bar{x} =x/\ell_{\mathrm{h}}$
where $\tau_\mathrm{h} = \ell_{\mathrm{h}} / \langle v \rangle$ is the characteristic hydrodynamic time scale.
In the new variables equation~\eqref{eq:ghd_boltzmann} takes the form
\begin{align}
   \partial_{\bar{t}}\rhop
            + \partial_{\bar{x}}\left(\overline{v}_{\rho} \,
                \rhop\right)
        = \frac{\deltaceii}{2} \partial_{\bar{x}}\left(
                \overline{\mathfrak{D}}_{\rho}
                    \partial_{\bar{x} }\rhop \right)
            + \frac{1}{\delta_\mathcal{I}} \, \bar{\mathcal{I}}[\rhop],
\end{align}
where  $\overline{v}_{\rho} = v_{\rho}/\langle v \rangle$, $\overline{\mathfrak{D}}_{\rho} =\mathfrak{D}_{\rho}/(\ell_{\mathrm{mic}} \, \langle v \rangle)$ and $\bar{\mathcal{I}} =\tau_{\mathcal{I}} \mathcal{I}$ are respectively velocity, diffusion operator and collision integral in the natural units.
In what follows we drop $\bar{\cdot}$ symbols from the notation and keep $\delta$ parameters. After truncating to a given order, we will restore the original units.
Since we assumed $\rhop$ to have the normal form~\eqref{eq:normal_solution} we can use the particle conservation law~\eqref{eq:ghd_conservation}
    to write the equation above without explicit time dependence:
\begin{align} \label{eq:ghd_ce}
\begin{split}
    \frac{\partial \rhop}{\partial \varrho}& G
        +  \partial_{x}\left(v_{\rho}
            \rhop\right) = \frac{\deltaceii}{2}\partial_{x}
            \left(\mathfrak{D}_{\rho}
                \partial_{x} \rhop\right)
        + \frac{1}{\delta_\mathcal{I}} \, \mathcal{I}[\rhop].
\end{split}
\end{align}
In the equation above we introduced function $G(x, \varrho) = - \partial_{x} j_\rho$, such that the continuity equation~\eqref{eq:ghd_conservation} becomes
\begin{equation} \label{continuity_ChE}
    \partial_{t} \varrho
    = G(x, \varrho).
\end{equation}
Solution of the ChE procedure will determine $G(x, \varrho)$ in terms of $\varrho$ thus closing the equation.

Next we proceed with expansion in small ChE parameters~\eqref{eq:ce_params_def}.
According to the usual scheme,  we expand the density $\rhop$ and $G$ in the two ChE parameters ~\cite{Resibois} and expand
\begin{align} \label{eq:expanded_quanties}
\begin{split}
    \rhop = \sum_{i, j = 0}^{\infty}
        \rhop^{(i, j)} \,
        \deltacei^i \,
        \deltaceii^j, \qquad
    G = \sum_{i, j = 0}^{\infty}
        G^{(i, j)} 
        \deltacei^i \, \deltaceii^j, \\
\end{split}
\end{align}
with $ G^{(i, j)} = - \int \dd\lambda \, \ulb{1} \, \partial_x \rhop^{(i, j)}$.
For notational clarity we introduce symbol $\bigO(\deltace^n)$ to mean all $\bigO(\deltacei^i \, \deltaceii^j)$ contributions such that $i + j = n$, $i \geq -1$, $j \geq 0$.
This notation becomes handy later.

In order to solve \eqref{eq:ghd_ce} perturbatively in ChE parameters we also have to expand $\mathcal{I}[\rhop]$, $v_{\rhop}$ and $\mathfrak{D}_{\rhop}$ in terms of ChE parameters.
Proceeding in analogy to~\eqref{eq:expanded_quanties} we denote expansion coefficients as $\mathcal{I}^{(i,j)}$, $v^{(i,j)}$ and $\mathfrak{D}^{(i,j)}$.
Since they are all functionals of $\rhop$ we need to relate each of them with $\rhop^{(i,j)}$.
This can be done by performing functional Taylor expansion of
$\mathcal{I}^{(i,j)}$, $v^{(i,j)}$ and $\mathfrak{D}^{(i,j)}$ around $\rhop^{(0,0)}$,
since $\rhop^{(0,0)}$ is the only factor of order unity in the expansion of $\rhop$.

Finally, and contrary to other quantities, we do not expand the conserved density $\varrho$. 
Since the only quantity of order unity in expansion of $\rhop$ is $\rhop^{(0,0)}$, for the theory to be consistent we require~\cite{Resibois}
\begin{align} \label{eq:n_rhopts_rel}
\begin{split}
    \varrho = \int \dd \lambda \, \rhop^{(0,0)}, \quad 
    0 = \int \dd \lambda \, \rhop^{(i,j)}
    \;\; \mathrm{for}  \;\; i,j \neq 0.
\end{split}
\end{align}
These are the constitutive relations that uniquely determine $\rhop^{(i,j)}$.

\subsection{Euler scale}

At the leading order $\bigO\left(\deltacei^{-1}\right)$, the GHD-Boltzmann equation~\eqref{eq:ghd_boltzmann} simplifies to
\begin{equation} \label{eq:ce_background_state}
    \mathcal{I}[\rhop^{(0,0)}] = 0.
\end{equation}
This condition fixes a state in each fluid cell to be stationary state of collision operator.
From now on, we will denote the stationary state $\rhop^{(0,0)}$ as $\st{\rhop}$.
In the present context, we assume that the stationary states are parametrized by their density. Therefore, equation~\eqref{eq:ce_background_state} has a family of solutions $\st{\rhop}(\lambda, \varrho)$ where $\varrho = \int {\rm d}\lambda \st{\rhop}$. 

To find the Euler scale equations we recall that the density $\varrho$ varies in space and the quasiparticle distribution is given by $\st{\rhop}(\lambda, \varrho(x,t))$. We can evaluate now the leading contribution to $G$ thus closing the continuity equation~\eqref{continuity_ChE}. We have
\begin{equation} \label{G_00}
    G^{(0, 0)} = - 
      \int \dd\lambda \, \ulb{1} \, \partial_x \rhop^{\rm st}
= - u \partial_x \varrho,
\end{equation}
where we used the chain rule and introduced
\begin{equation} \label{eq:xi_def}
    u = \int \dd \lambda\,\ulb{1} \frac{\partial \st{\rhop}}{\partial \varrho}.
\end{equation}
Changing variables to the original, dimensionful, ones we arrive at
    the Euler-scale evolution equation
\begin{equation} \label{eq:euler_scale_evolution}
   \partial_t \varrho
        + u(\varrho) \partial_x{\varrho} = 0.
\end{equation}

Let us now gain some intuition on the advective coefficient $u$.
First of all, it can be shown that under the Galilean boost with velocity $w$, it transforms indeed as velocity: $u \rightarrow u + w$ (see Appendix~\ref{app:galilean}). This renders the Euler-scale equation Galilean invariant as it should be. 

Furthermore, for parity-symmetric states (this includes thermal states) $u = 0$. This follows from the observations that the derivative can be taken in front of the integral and that the first moment of a symmetric distribution is $0$. A Galilean boost makes the first moment finite but density independent. Therefore, for standard collision integral with only thermal stationary term there is no dynamics at the Euler scale. A result known from the standard kinetic theory. However, for non-thermal stationary states $u$ can be non-zero. This requires the stationary state to be asymmetric in $\lambda$ and this asymmetry cannot be removable by a boost. 

\subsection{Diffusive scale}

Let us now turn to finding the first-order contribution to~\eqref{eq:ghd_ce}. There will be two contributions at order $\delta^0$ from the collision integral
\begin{equation}
    \frac{1}{\delta_\mathcal{I}} \mathcal{I}[\rhop] = \st{\Gamma} \cdot \left[\rhop^{(1,0)} + \frac{\delta_\mathcal{D}}{\delta_\mathcal{I}} \rhop^{(0,1)}\right] + \mathcal{O}(\delta),
\end{equation}
where we introduced linear operator
\begin{equation} \label{Gamma}
    \st{\Gamma} = \frac{\delta \mathcal{I}}{\delta \rhop}\Big|_{\st{\rhop}}.
\end{equation}
At the same time the GHD diffusion does not contribute at this order. This gives then two equations
\begin{align}
    \st{\Gamma} \cdot \rhop^{(1,0)} &= \frac{\partial \st{\rho}}{\partial \varrho} G^{(0,0)} + \frac{\partial \left(\st{v} \st{\rho}\right)}{\partial x} , \\
    \st{\Gamma} \cdot \rhop^{(0,1)} &= 0.
\end{align}
The second equation implies that $\rho^{(0,1)}$ is in the kernel of $\st{\Gamma}$. This implies that $\rho^{(0,1)} = 0$, due to constitutive relations. We are left with the equation for $\rho^{(1,0)}$. We substitute for $G^{(0,0)}$ and use the chain rule in the last term to obtain
\begin{equation} \label{int_eq_ChE_0}
        \st{\Gamma} \cdot \rho^{(1,0)} = \left(\st{\mathcal{A}} - u\right) \cdot\frac{\partial \st{\rhop}}{\partial \varrho} \frac{\partial \varrho}{\partial x},
\end{equation}
where we used that the hydrodynamic matrix $\mathcal{A} \equiv \mathcal{B} \mathcal{C}^{-1}$ obeys $\delta (v \rhop) = \mathcal{A} \cdot \delta \rho_p$~\cite{drudeweight}.
The integral equation~\eqref{int_eq_ChE_0} is simplified by the following ansatz
\begin{equation}
    \rhop^{(1,0)}(\lambda, \varrho(x)) = a(\lambda, \varrho(x)) \partial_x \varrho.
\end{equation}
The resulting equation, free of explicit space-time dependence, is then
\begin{equation} \label{int_eq_ChE}
        \st{\Gamma} \cdot a = \left(\st{\mathcal{A}} - u\right) \cdot \frac{\partial \st{\rhop}}{\partial \varrho}.
\end{equation}
The constitutive relation~\eqref{eq:n_rhopts_rel} additionally implies
\begin{equation} \label{constitutive_relation}
    \int {\rm d}\lambda\, a(\lambda, \varrho) = 0.
\end{equation}

The resulting contribution to the hydrodynamic equation is $G^{(1,0)} = \partial_x \left( \mathcal{D}(\varrho) \partial_x \varrho\right)$
where we defined the diffusion coefficient
\begin{equation} \label{CE_diffusion}
    \mathcal{D}(\varrho) = - \int {\rm d}\lambda h_1(\lambda) a(\lambda, \varrho).
\end{equation}
By combining the Euler scale equation with the contribution from the diffusive scale, we finally obtain the Burgers~\eqref{Burgers_intro}.
The diffusion coefficient $\mathcal{D}(\varrho)$ is invariant under the Galilean transformations, see Appendix~\ref{app:galilean}. Note that there is no contribution from the diffusion present in the GHD as $G^{(0,1)} = 0$. We also mention that the diffusion coefficient in general scales as an inverse of characteristic timescale of $\mathcal{I}$ meaning that it is parametrically big in the strength of integrability breaking. This concludes the derivation of the hydrodynamic equation for density.

\section{Applications} \label{sec:applications}
In the following section we apply the ChE formalism to coupled Bose gases, stochastic velocity swap models and to the so-called relaxation time approximation of collision operator. 

\subsection{Relaxation time for two tubes system} \label{two_tubes_system}

We turn now to a more detailed picture of the collision integral which arises in a quasi-1D systems of two 1D gases that are coupled with a density-density interaction. In the density imbalance limit, as discussed in Section~\ref{sec:athermal_states}, the collision integral $\mathcal{I}[\rho_{\rm p}]$ in the minority tube, assumed to be tube 1, is given by the Master equation~\eqref{master}.
The dynamic structure factor of the bath tube is given by~\cite{Panfil2023}
\begin{align} \label{DynamicStructureFactor}
\begin{split}
    S^{(2)}_0&(k, \omega) = (2 \pi)^2 \int \dd \lambda \dd \mu \,
        \rhop^{(2)}(\lambda) \rhoh^{(2)}(\mu)  \\
    & \times |F_2(\lambda, \mu)|^2 \delta(k - k_2(\lambda, \mu)) \delta(\omega - \omega_2(\lambda, \mu)).
\end{split}
\end{align}
In the density imbalance limit evolution of the environmental tube is frozen i.e., $\partial_t \rhop^{(2)} = \mathcal{I}_{21}[\rhop^{(2)}, \rhop^{(1)}] \approx 0$.
Therefore we are free to choose $\rhop^{(2)}$ to be any stationary state.
We will denote the state in the environmental tube as $\rhop^{(b)}$ and will refer to the particle density in the evolving tube $\rhop^{(1)}$ simply as $\rhop$.
Let us assume that (homogenous) state of the system tube is close to a stationary state
\begin{equation}
    \rhop = \st{\rhop} + \delta \rhop,
\end{equation}
where $\st{\rhop}$ satisfies $\mathcal{I}_{12}[\st{\rhop}] = 0$.
In such setup, evolution of the system tube can be approximated as
\begin{equation}
    \frac{\partial \left(\delta \rhop\right)}{\partial t} = \st{\Gamma} \cdot \delta \rhop,
\end{equation}
where
\begin{equation} \label{LinearizedFullDef}
    \st{\Gamma} = \frac{\delta \mathcal{I}_{12}[\rhop, \rhop^{(b)}]}{\delta \rhop}\Big|_{\rhop = \st{\rhop}}.
\end{equation}
Detailed computation of $\st{\Gamma}$ is presented in the appendix~\ref{app:linearization}.
Given that $\mathcal{I}_{12}$ has only one collision invariant, operator $\st{\Gamma}$
    has exactly one left null vector, corresponding to the conservation of the number of particles, see~\eqref{eq:boltzmann_conservation}.
Denoting eigenvalues as $\gamma_i$ we have $0 = \gamma_0 > \gamma_1 \geq \gamma_2 \geq \ldots$.
Except for the non-decaying zeroth mode, each eigenmode of $\st{\Gamma}$ will decay with the characteristic time $1/|\gamma_i|$.
The characteristic relaxation time can be identified with a gap between zero mode and smallest non-zero eigenvalue of $\st{\Gamma}$~\cite{Lopez-Piqueres2021}.
\begin{equation}
    \taurta = |\gamma_1^{-1}|.
\end{equation}
Numerical estimation of the relaxation time is discussed in the Appendix~\ref{app:numerical}. This gives a conservative thermalisation scale and observables that do not couple to the slowest decaying charge will thermalise at faster rate.

\subsection{Relaxation Time Approximation}

Given the complex nature of the collision integral, to be able to say more about the evolution without restricting to the specific system we need to make some approximation.
One of the simplest approximation to apply is the RTA~\cite{Lopez-Piqueres2021}.
It amounts to the following approximation of collision integral
\begin{equation} \label{eq:rta_collision_integral}
    \mathcal{I}[\rhop]
        \approx \taurta^{-1} \, \left(\st{\rhop}[\rhop] - \rhop\right),
\end{equation}
where multiplication by $\taurta^{-1}$ enforces relaxation at proper time scale.
We only consider the case when the relaxation time is independent of $\varrho$.
In the equation above $\st{\rhop}$ is a stationary state towards which $\rhop$ evolves.
If the only non-trivially conserved charge is the total number of particle then $\st{\rhop}$
    is related to $\rhop$ via $\int \dd \lambda \, \st{\rhop} = \int \dd \lambda \, \rhop$.
This relation makes $\st{\rhop}[\rhop]$ highly non-linear functional of $\rhop$~\cite{Lopez-Piqueres2021}.
Our ultimate goal in this section is to compute the diffusion coefficient~\eqref{CE_diffusion} for the collision 
    integral in the relaxation time approximation.
First thing to do is to compute linearized collision integral~\eqref{Gamma}:
\begin{equation}
    \st{\Gamma} = \frac{1}{\taurta} \left(\frac{\partial \st{\rhop}}{\partial \varrho} h_0 - \mathbf{1}\right),
\end{equation}
where $\mathbf{1}$ is the identity operator represented by the Dirac $\delta$-function. Because of the constitutive relation~\eqref{constitutive_relation}, action of $\Gamma$ on $a$ is especially simple
\begin{equation}
    \st{\Gamma} \cdot a = - \frac{1}{\taurta} a,
\end{equation}
since
\begin{equation}
     \frac{\partial \st{\rhop}(\lambda)}{\partial \varrho}\int {\rm d}\mu\, h_0(\mu) a(\mu) = 0.
\end{equation}
The integral equation is now simple to solve,
\begin{equation}
    a = \taurta \left(\st{\mathcal{A}} - u\right) \cdot \frac{\partial \st{\rhop}}{\partial \varrho}.
\end{equation}
By direct computations one can check that the constitutive relation~\eqref{constitutive_relation} is fulfilled. The resulting diffusion coefficient is
\begin{equation}
    \mathcal{D}(\varrho) = \taurta \left(\int {\rm d}\lambda h_1(\lambda) \st{\mathcal{A}} \cdot \frac{\partial \st{\rhop}}{\partial \varrho} - u^2\right).
\end{equation}
This formula can be rewritten in terms of the momentum current
\begin{equation}
    j_1 = \int {\rm d}\lambda\, h_1(\lambda) v(\lambda) \st{\rhop}(\lambda).
\end{equation}
Computing the derivative of $j_1$ with respect to the density we find
\begin{equation}
    \frac{\partial j_1}{\partial \varrho} = \int {\rm d}\lambda\, h_1(\lambda) \frac{\partial(v(\lambda) \st{\rhop})}{\partial \varrho} = \int {\rm d}\lambda\, h_1(\lambda) \st{\mathcal{A}} \frac{\partial\st{\rhop}}{\partial \varrho}.
\end{equation}
Hence the diffusion coefficient is
\begin{equation}
    \mathcal{D}_{\rm RTA}= \taurta \left(\frac{\partial j_1}{\partial \varrho} - u^2 \right).
\end{equation}

In a thermal state $u=0$ and the momentum current is equal to the thermodynamic pressure. The derivative of pressure is computed at fixed temperature which relates it to the isothermal compressibility $\kappa_T$ such that
\begin{equation} \label{D_RTA_th}
    \mathcal{D}_{\rm RTA}^{\rm th} = \frac{\taurta}{\varrho \kappa_T}, \qquad\frac{1}{\kappa_T} = \varrho \left(\frac{\partial P}{\partial \varrho}\right)_T,
\end{equation}
where $P$ is the 1D pressure.
The diffusion coefficient $\mathcal{D}_{\rm RTA}^{\rm th}$ is positive since $\kappa_T>0$ reflecting the stability of thermal equilibrium.

\begin{figure}[t]
\includegraphics[scale=0.50]{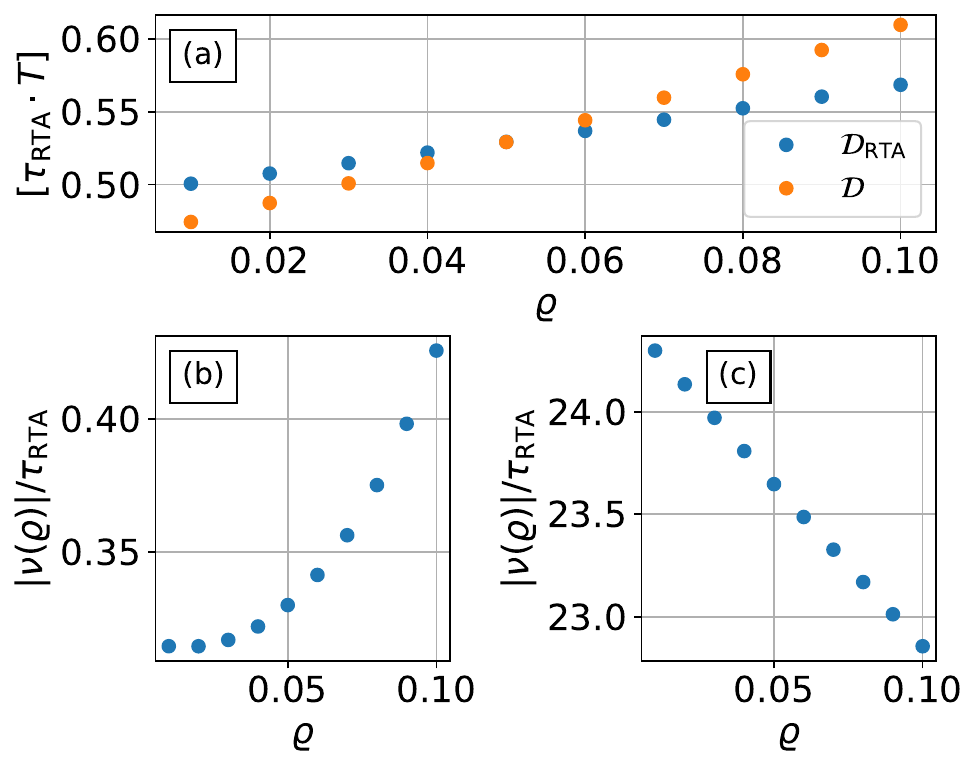}
\caption{Transport coefficients in the coupled LL models in the Tonks-Girardeau limit. Panel (a): comparison of diffusion coefficient stemming from RTA with non-approximated diffusion for the two-tubes system in a thermal state ($T = 4.0$). In the panels (b) and (c) we show the Burgers parameter $\nu = \mathcal{D}_{\rm{RTA}} / u'$ for a system in asymmetric Bragg-split state~\eqref{AsymmetricBraggSplitState}. In (b) we present a system in a small temperature ($T = 1.0$), with asymmetry $b=9$ with the peak separation $\lambda_B=3.9$ chosen in a way so that they overlap over a small region in $\lambda$. For (c),  we increased the temperature ($T = 10.0$), simultaneously increasing peak separation to $\lambda_B=8.0$ with the same $b=9$.}
    \label{TwoTubesPlot}
\end{figure}

In the upper part of Fig.~\ref{TwoTubesPlot} we present comparison of the diffusion coefficient computed in the RTA with the diffusion coefficient computed using the full collision integral for the two (Lieb--Liniger) tubes system~\footnote{The inter-tube interaction potential was chosen to be a Gaussian potential $V(x) = V_0 e^{-\left(x/x_r\right)^2} / (\sqrt{\pi} \, x_r)$ with $x_r = 2$. Constant $V_0$ can be expressed as $\tauib^{-1} = 2 A_0^2 m / \hbar$~\cite{Lebek2024}. For the purpose of numerical computations we set $m = 1$. Characteristic relaxation time is extracted from spectrum of the linearized collision integral in the limit $\varrho \to 0$.}.
For more detailed description of the numerical procedure we refer to the Appendix~\ref{app:numerical}.
For the analyzed system we observe that RTA gives a good approximation of the diffusion coefficient.
In the lower panel figure~\ref{TwoTubesPlot} we present numerically-computed $\nu$ for the bath tube in an asymmetric Bragg-split state, i.e., state of the form
\begin{equation} \label{AsymmetricBraggSplitState}
    \rhop^{(b)}(\lambda) \propto 
        \thm{\rhop}(\lambda + \lambda_B)+b \thm{\rhop}(\lambda -\lambda_B),
\end{equation}
where $\thm{\rhop}$ is a Gibbs thermal state, $b \in [0, 1]$ is an asymmetry parameter and $\lambda_B$ fixes the peak separation.
Numerically computing $u'$, we find that its value varies over several orders of magnitude, depending on the environmental tube's state.
We find that $u'$ attains the largest values in the regime of low temperatures, with $\lambda_b$ being slightly smaller than $({\rm width \; of \;}\thm{\rhop}) / 2$, meaning that left and right peaks minimally overlap with each other.

Let us now present a qualitative way of deriving the diffusion coefficient for a classical thermal case.
Suppose we have a viscous stationary fluid. A slowly moving particle experiences then a Stokes drag
$F_d = -\zeta \, u$
where $u$ is the particle's velocity and $\zeta$ is friction.
Under such assumption the total force acting on the particle is $F = F_p + F_d$, where
    $F_p = - a \, \partial_x P$ is a force exerted by the pressure gradient on a particle of size (length) $a$.
The net force $F$ will cause the particle to accelerate towards the terminal velocity fixed by the condition $F = 0$.
Thus, assuming the particle to reach the terminal velocity quickly, we get a relation between
    velocity of the particle and the pressure gradient
\begin{equation}
    u = - \frac{{\rm d}a}{\zeta} \frac{\partial P}{\partial x}.
\end{equation}
Conservation of the total number of particles leads to the  continuity equation $ \frac{\partial \varrho}{\partial t} + \partial_x(\varrho u) = 0$. Using the relation between $u$ and the pressure gradient and assuming that each fluid cell is in a thermal state with fixed temperature and locally varying density we obtain
\begin{equation}
    \frac{\partial \varrho}{\partial t} - \frac{\partial}{\partial x}\left(\frac{m}{\zeta}\frac{\partial P}{\partial \varrho} \frac{\partial \varrho}{\partial x}\right) = 0,
\end{equation}
from which the diffusion coefficient can be read off. Comparing with the result of the ChE theory within the RTA we find the relation between the friction and the relaxation time given by $\taurta = \frac{m}{\zeta}$.
This provides a way to estimate the relaxation time from the kinematics of a single particle.

\subsection{Stochastic velocity swap models}

In this section we analytically compute diffusion coefficient $\mathcal{D}(\varrho)$ along with $u(\varrho)$ for the velocity swap model of classical and fermionic particles. We can treat both cases at the same time by introducing $\chi$ which is $\chi=0$ for classical particles and $\chi=1$ for fermions. We begin by introducing pseudoenergy functions
\begin{equation} \label{EpsDefCoupledToClassical}
    \epsilon = \log\left(\frac{1 - \chi n}{n}\right), \qquad \epsilon^{(b)} = -\log(n^{(b)}).
\end{equation}
First, let us find the stationary states. To this end we write the collision integrals~\eqref{eq:swap1} and~\eqref{eq:swap2} in a convenient form
\begin{align}
\begin{split}
        \mathcal{I}[\rhop](\lambda) = \gamma &\int \dd \lambda' (1 - \chi n(\lambda')) \rhop^{(b)}(\lambda') \rhop(\lambda) \\
        &\times \left(e^{\epsilon(\lambda) - \epsilon(\lambda') - \left(\epsilon^{(b)}(\lambda) - \epsilon^{(b)}(\lambda')\right)} - 1\right).
\end{split}
\end{align}
The condition fixing the stationary state gives a simple relation between pseudo-energies
\begin{equation} \label{PseudoEnergyRelationToyModel}
    \st{\epsilon}(\lambda) = \epsilon^{(b)}(\lambda) + \log \tilde{A},
\end{equation} where $\tilde{A}$ is some positive constant.
From~\eqref{EpsDefCoupledToClassical} we recover the stationary states~\eqref{eq:stat_ferm} and~\eqref{eq:Master_stat}. The constant $A=\frac{\tilde{A}}{2 \pi}$ is fixed by the condition $\varrho = \int \dd \lambda\, \st{\rhop}(\lambda)$.
It follows that for a fixed state $\rhop^{(b)}$, $A$ is a function of $\varrho$ only and so
$\frac{\partial \st{\rhop}(\lambda, \varrho)}{\partial \varrho} = \frac{\partial \st{\rhop}(\lambda, A)}{\partial A} \frac{\dd A}{\dd \varrho}$. By a direct computation we get
\begin{equation}
    \frac{\partial \st{\rhop}(\lambda, A)}{\partial A}=-A^{-1} \st{\rhop}(\lambda) (1 - \chi\st{n}(\lambda)).
\end{equation}
Similarly $\frac{\dd \varrho}{\dd A} = -A^{-1} \hst{h_0}{h_0}$, where we defined 
\begin{equation}
    \hst{\phi}{\psi} := \int \dd \lambda \st{\rhop}(\lambda) (1 - \chi\st{n}(\lambda)) \, \phi(\lambda) \psi(\lambda).
\end{equation}
Therefore we get
\begin{equation}
    \frac{\partial \st{\rhop}(\lambda, \varrho)}{\partial \varrho} = \frac{\st{\rhop}(\lambda) (1 - \chi\st{n}(\lambda))}{\hst{h_0}{h_0}}.
\end{equation}
It follows that for the analyzed system $u$ is given by
\begin{equation}
    u(\varrho) = \frac{\hst{h_1}{h_0}}{\hst{h_0}{h_0}}.
\end{equation}
In similar manner we find
\begin{equation}
    u'(\varrho) = -2 \chi \frac{\hst{h_1 - u}{\st{n}}}{\hst{h_0}{h_0}^2}.
\end{equation}

Now we turn to computing the diffusion coefficient. To do this we need to solve the linear integral equation~\eqref{int_eq_ChE}.
Before proceeding let us note that for the considered system $\st{v}(\lambda) = \lambda$ and we can identify $\mathcal{A} \equiv \lambda$.
Thus the equation we need to solve is
\begin{equation}
    \frac{\delta \mathcal{I}}{\delta \rhop}\Big|_{\st{\rhop}} \cdot a = \frac{(\lambda - u) \st{\rhop}(\lambda) (1 - \chi\st{n}(\lambda))}{\hst{h_0}{h_0}}.
\end{equation}
We proceed by writing
\begin{equation}
\begin{aligned}
    \frac{\delta \mathcal{I}}{\delta \epsilon}\Big|_{\st{\rhop}} \cdot \delta \epsilon = \gamma \int \dd \lambda' &(1 - \chi \st{n}(\lambda')) \rhop^{(b)}(\lambda') \rhop(\lambda) \\
        &\times \left[\delta \epsilon(\lambda) - \delta \epsilon(\lambda')\right],
\end{aligned}
\end{equation}
where we used that for a stationary state equation~\eqref{PseudoEnergyRelationToyModel} is satisfied.
From the definition~\eqref{EpsDefCoupledToClassical} it follows that $\delta \rhop = - \rhop (1 - \chi n) \delta \epsilon$, which can be checked by a direct calculation.
Introducing
\begin{equation}
    b(\lambda) = -\gamma \frac{\hst{h_0}{h_0}\, a(\lambda)}{\st{\rhop}(\lambda) (1 - \chi\st{n}(\lambda))}
\end{equation}
the integral equation becomes
\begin{equation} \label{FermiClassicalIntEqn}
    \int \dd \lambda' (1 - \chi \st{n}(\lambda')) \rhop^{(b)}(\lambda')\left[b(\lambda) - b(\lambda')\right] = (\lambda - u) (1 - \chi \st{n}).
\end{equation}
To solve the equation we differentiate both sides over $\lambda$
\begin{equation}
    b'(\lambda) = \frac{1}{\hst{h_0}{\rhop^{(b)}/\st{\rhop}}}\left(\left(\lambda - u\right)\left(1 - \chi \st{n}(\lambda)\right)\right)',
\end{equation}
and so
\begin{equation}
    b(\lambda) = \frac{\left(\lambda - u\right)\left(1 - \chi \st{n}(\lambda)\right) - C}{\hst{h_0}{\rhop^{(b)}/\st{\rhop}}},
\end{equation}
with $C$ being some constant.
\begin{figure}[t]
\includegraphics[scale=0.47]{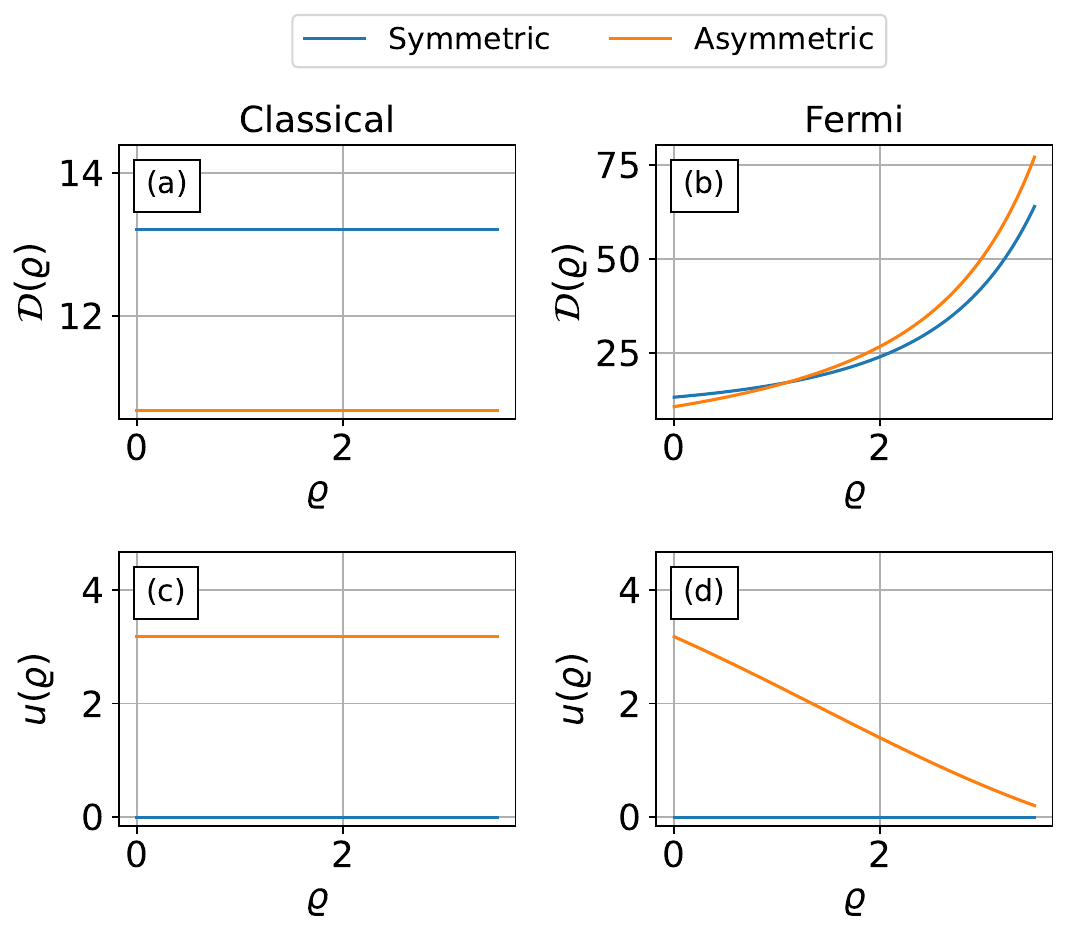}
\caption{Diffusion (panels (a) and (b)) and advective (panels (c),(d)) coefficients as a function of density $\varrho$ for the stochastic velocity swap models coupled to the classical bath. The bath is in the same state as in  Fig.~\ref{MolecularDynamicsPlots} and the collsion rate $\gamma=0.08$ is the same as well. Note that fermionic statstics make both diffusion and advective coefficients density-dependent. This has important consequences for the dynamics as visible in Fig.~\ref{MolecularDynamicsPlots}(d).}
\label{fig:master_transport_coeff}
\end{figure}

\begin{figure*}[t]
\includegraphics[scale=0.58]{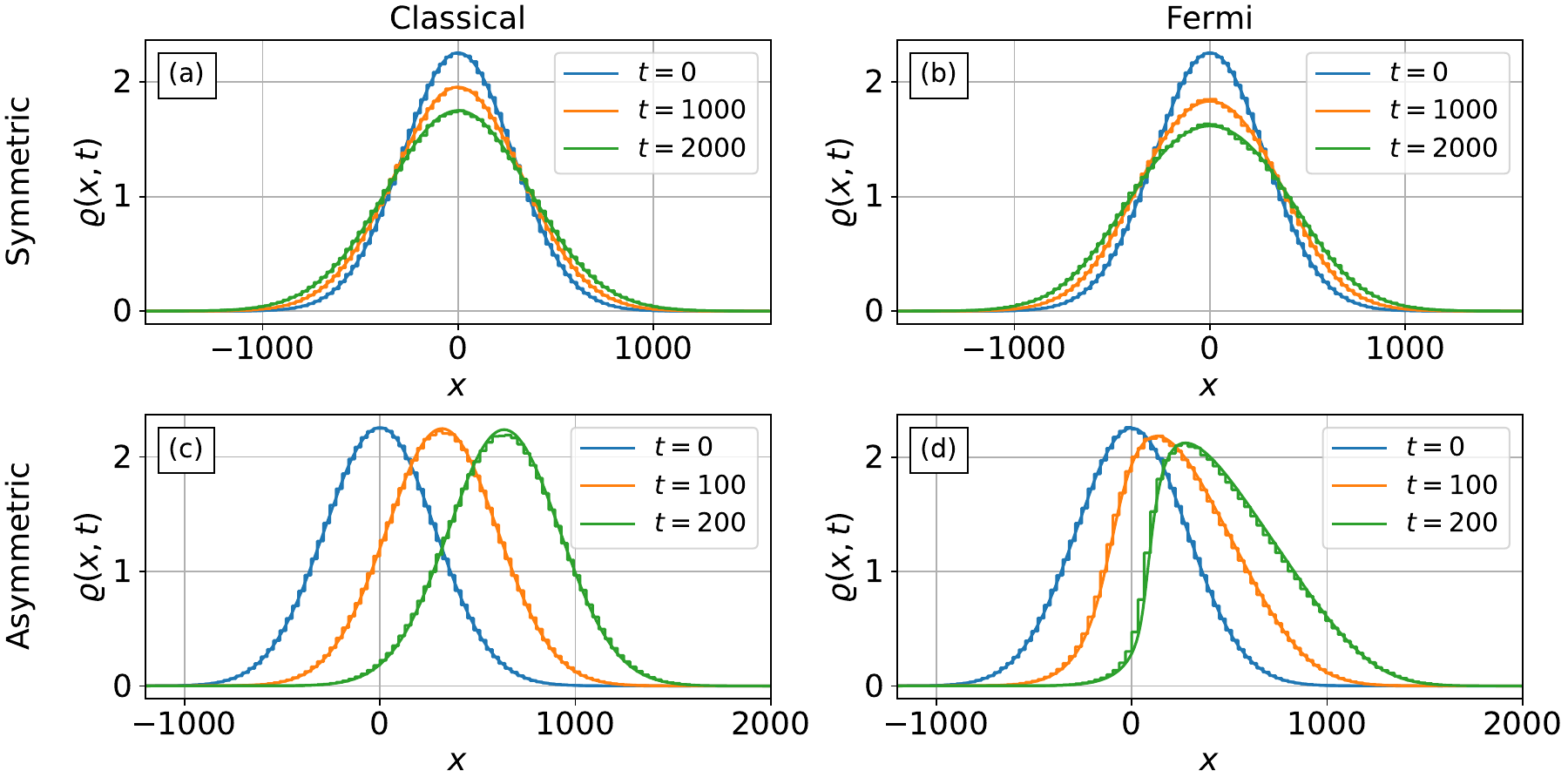}
\caption{Molecular dynamics simulations for velocity swap models ($\gamma = 0.08$, $\varrho^{(b)} = 50.0$, $N=1600$) with initial states given by \eqref{eq:initial_states}. We observe an excellent agreement between numerical simulations (histograms) and hydrodynamic equation \eqref{Burgers_intro} derived through ChE method (solid lines) for all cases. For symmetric states in bath system, cf. panels (a,b), both systems are described by diffusion equation~\eqref{eq:diff_eq}. Upon introducing the asymmetry, the classical system is described by boosted diffusion equation, see panel (c). On the other hand, the fermionic model exhibits Burgers equation~\eqref{Burgers_intro} dynamics, as visible in the dynamically generated asymmetry in the initially Gaussian shape, cf. panel (d). Results from microscopic simulations were averaged over 1000 realizations.}
\label{MolecularDynamicsPlots}
\end{figure*}

To fix the constant we use the relation~\eqref{constitutive_relation}, which gives
\begin{equation}
     b(\lambda) = \frac{\left(\lambda - u\right)\left(1 - \chi \st{n}(\lambda)\right) - \frac{\hst{\lambda - u}{1 - \chi \st{n}}}{\hst{h_0}{h_0}}}{\hst{h_0}{\rhop^{(b)}/\st{\rhop}}}.
\end{equation}
Let us now verify that the obtained solution indeed solves the integral equation.
Plugging it into~\eqref{FermiClassicalIntEqn} we find that the equation has a solution iff
\begin{equation}
    \int \dd \lambda' (1 - \chi \st{n}(\lambda')) \rhop^{(b)}(\lambda') \left(\lambda' - u\right)\left(1 - \chi \st{n}(\lambda')\right) = 0.
\end{equation}
To see why the equation above is always satisfied, note that for a stationary state
\begin{equation}
    (1 - \chi \st{n}(\lambda')) \rhop^{(b)}(\lambda') = A \st{n}(\lambda'),
\end{equation}
which follows from the definition~\eqref{EpsDefCoupledToClassical} and~\eqref{PseudoEnergyRelationToyModel}.
Thus the solvability condition is equivalent to $\hst{h_1 - u}{h_0} = 0$, which coincides with the definition of $u$, hence the solution we have found is the correct solution of~\eqref{FermiClassicalIntEqn}.
Having solved for $b(\lambda)$, we compute the diffusion coefficient using~\eqref{CE_diffusion} which, after some algebraic manipulations, gives us
\begin{equation}
    \mathcal{D}(\varrho) = \frac{1}{\gamma}\frac{\left(\left(h_1 - u\right)^{2}\Big|1 - \chi \st{n}\right)_{\rm st}}{\hst{h_0}{h_0} \hst{h_0}{\rhop^{(b)}/\st{\rhop}}}.
\end{equation}
For the classical particles $\chi = 0$ and the relation between the bath state and a stationary state is especially simple and given by~\eqref{eq:Master_stat}. The convection velocity function $u(\varrho)$ simplifies to
\begin{equation}
    u = \frac{1}{\varrho^{(b)}} \int \dd \lambda \lambda \rhop^{(b)}(\lambda) = \rm{const},
\end{equation}
and the diffusion coefficient is
\begin{equation}
    \mathcal{D} = \frac{1}{\gamma \left(\varrho^{(b)}\right)^2} \int \dd \lambda (\lambda - u)^2 \rhop^{(b)}(\lambda) = \rm{const}.
\end{equation}
Hence, performing a boost to remove the convective term, the general Burgers-diffusion~\eqref{Burgers_intro} equation reduces to the ordinary linear diffusion equation
\begin{equation}
    \partial_t \varrho  = \mathcal{D} \partial_x^2 \varrho.
\end{equation}
For $\rhop^{(b)}$ being a boosted thermal state with pseudo energy $\epsilon^{(b)} =  \beta \, \mu^{(b)} + \kappa \, \ulb{1}(\lambda) + \beta \, \ulb{2}(\lambda)$, the diffusion coefficient reads
\begin{equation}
    \mathcal{D} = \frac{T}{\gamma \varrho^{(b)}}.
\end{equation}
In Fig.~\ref{fig:master_transport_coeff} we present the transport coefficient computed for classical and fermionic particles for the $\rho^{(b)}$ being an asymmetric Bragg split state.

\section{Numerical simulations} \label{sec:numerics}
We consider now molecular dynamics of the stochastic velocity swap models to check the hydrodynamic theory for both classical and fermionic particles. Both cases can be simulated efficiently and on the level of kinetic equation description, they are given by \eqref{eq:swap1} and \eqref{eq:swap2}.

In the simulation, particles are initialized such that the initial distribution of velocities is given by
\begin{equation}\label{eq:initial_states}
    \rho_{\rm p}^{(b)} \propto e^{-(\lambda+\lambda_B)^2/\sigma^2}+be^{-(\lambda-\lambda_B)^2/\sigma^2}
\end{equation}
with $\sigma=6$, $\lambda_B=5.9$ and $b=1$ for symmetric states and $b=0.3$ for asymmetric states. The particles propagate freely and occasionally undergo two-body collisions with the bath. A collision amounts to velocity swap between system and a bath. When particles are classical and collisions happen at a constant rate the corresponding Master equation is precisely \eqref{eq:swap1}. Since the rate does not depend on the state of the system, such collisions can be efficiently implemented with standard Gillespie algorithm~\cite{Gillespie1977}.
The more complicated case is the simulation of fermionic kinetic equation given by \eqref{eq:swap2}. There, we partition the system into fluid cells and propose an algorithm, which effectively incorporates the Pauli principle into the collisions. Algorithms for both cases are discussed in Appendix~\ref{app:algorithm}.

The results are presented in Fig.~\ref{MolecularDynamicsPlots}. A very good agreement is found between molecular simulations and hydrodynamic equation~\eqref{Burgers_intro}. Depending on symmetry of the state (or lack thereof) and particle statistics, we find that numerical results excellently agree with diffusion equation~\eqref{eq:diff_eq} and with Burgers equation~\eqref{Burgers_intro} for fermionic model with parity-asymmetric state.

Burgers equation in 1D is well-known for its breakdown of diffusive behavior in hydrodynamic two-point functions of the density field~\cite{Spohn1985} close to homogenous equilibrium state. From this perspective, it is interesting to note that (finite) diffusion constant computed from kinetic theory is relevant for description of the numerics in our highly inhomogenous setting.

\section{Summary and discussion}

In this work, we have considered a canonical problem of diffusion in which a minority species undergoes an irreversible motion due to coupling to the majority species, whose dynamics is not monitored and which, as is customary to assume, is in a spatially homogenous state. Due to the conservation of particle number of the minority species, its density obeys a continuity equation. In the standard situation of classical particles this continuity equation leads to a (possibly non-linear) diffusion equation. In this work we have identified the conditions under which the density follows instead the Burgers equation. 

The two sufficient conditions are: a) athermal and parity-asymmetric state of the majority species (bath) and b) quantum statistics of minority species (system).

We have demonstrated that these two conditions are fulfilled for effective dynamics of nearly-integrable coupled gases. In such scenario, the integrability naturally makes athermal states possible. We have provided a kinetic explanation for appearance of such athermal states through limitations of two-body scattering in 1D. In our approach, we assume a large density imbalance between the two gases allowing for a separation of timescales between the slow dynamics in the dense tube (bath) and fast dynamics in the dilute tube (system). Starting from the GHD-Boltzmann description, we have derived an effective hydrodynamic equation for the density of the system particles (minority species). We have shown that the resulting equations become a standard diffusion equation when the bath is in a thermal state. Instead, for athermal state of a bath we obtain Burgers equation. Our method provides also expressions for transport coefficients: advective coefficient $u(\varrho)$ and diffusion constant $\mathcal{D}(\varrho)$.

To unravel the essential ingredients leading to the Burgers equation we have considered also a simple stochastic model of two-particle scatterings between system and the bath. We observed an excellent agreement between microscopic numerical simulations and hydrodynamics derived from kinetic equation describing such random scattering events.

Lastly, let us discuss the features of the hydrodynamic density-density correlation function, which is a natural object appearing in the literature on 1D Burgers equation~\cite{Spohn1985}. This is because under the presence of ballistic mode with a nonlinear current, the correlation function is expected to exhibit a breakdown of diffusive behavior at the largest scales~\cite{Spohn2014}. Close to the integrability, we generically expect large diffusion constants, which push away the anomalous broadening to very large space-time scales. In this sense, we expect a scenario similar to the one analyzed in the context of Navier-Stokes equations~\cite{chapmanenskog}. We also note that one has to be careful when looking at asymptotically large scales in models with (late-time) thermalization channels. There are three-particle (and higher) collision processes, which provide a mechanism for thermalization at late times, assuming that the density of the bath is finite. This leads ultimately to the breakdown of Burgers equation. We plan to investigate the interesting problem of two-point functions using the models studied here in the future.

\begin{acknowledgments}
M.{\L}.~and M.P.~acknowledge support from the National Science Centre, Poland, under the  OPUS grant 2022/47/B/ST2/03334.
\end{acknowledgments}

\appendix

\section{Grazing limit of the collision integral~\eqref{eq:Master_general}}\label{app:grazing}
In this Appendix we formalize the grazing limit introduced in Sec.~\ref{sec:athermal_states}. Our starting point is collision integral ~\eqref{eq:Master_general}. We introduce new variables under the integral
\begin{equation}
    \alpha=\lambda-\lambda', \qquad \gamma=\frac{\lambda+\lambda'}{2}
\end{equation}
and similarly we define $\alpha_b$ and $\gamma_b$. Note that the parameters $\alpha,\alpha_b$ are nothing else but changes of momenta of system (bath) particles that take part in a collision process $\lambda'+\lambda_b' \to \lambda+\lambda_b$. The collision rate becomes
\begin{equation}\label{eq:ratealphagamma}
    \mathcal{W} \left(\gamma+\frac{\alpha}{2},\gamma_b+\frac{\alpha_b}{2};\gamma-\frac{\alpha}{2},\gamma_b-\frac{\alpha_b}{2}\right)
\end{equation}
and the grazing limit is easiest to formulate when the expression above is viewed as a function of $\alpha$ and $\alpha_b$. We essentially assume that it is localized in the region where both $\alpha,\alpha_b$ are small as compared to typical range of variation of states $\rhop$ and $\rhop^{(b)}$. This allows for performing expansions in $\alpha,\alpha_b$ in other terms, that enter the integral. For convenience, we rewrite the collision integral into the form
\begin{equation}
\begin{aligned}
    \mathcal{I}[\rhop](\mu)&= \int \dd \lambda \dd \lambda' \dd \lambda_b \dd \lambda_b'\delta(\mu-\lambda) \mathcal{W}(\lambda,\lambda_b;\lambda',\lambda_b')  \times\\
    &\delta(k) \delta(\omega)\rhop(\lambda')\rhop^{(b)}(\lambda) \times \\
    &\left(\exp\left[\epsilon(\lambda)+\epsilon_b(\lambda_b)-\epsilon(\lambda')-\epsilon_b(\lambda_b')\right]-1\right),
\end{aligned}
\end{equation}
For instance, the momentum conservation becomes
\begin{equation}
    k'(\gamma)\alpha+k_b'(\gamma_b)\alpha_b=0
\end{equation}
and similarly for energy and expression with pseudoenergies. This yields
\begin{equation}
    \delta(k)\delta(\omega)= \frac{\delta\left(v(\gamma)-v_b(\gamma_b)\right)\delta\left(\alpha_b+\alpha\frac{k'(\gamma)}{k'_b(\gamma_b)}\right)}{|\alpha| k'(\gamma) k_b'(\gamma_b)}.
\end{equation}
We also expand $\delta(\mu-\gamma-\frac{\alpha}{2}) = \delta(\mu-\gamma)- \partial_\mu \delta(\mu-\gamma) \frac{\alpha}{2}+ \ldots$. Moreover, we assume that the rate~\eqref{eq:ratealphagamma} is symmetric function of $\alpha$ and thus the first term in the expansion of Dirac delta vanishes. We define
\begin{equation}
\begin{aligned}
    &\tilde{\mathcal{W}}(\gamma,\gamma_b) = \frac{1}{k'(\gamma)k'(\gamma_b)}\int \dd \alpha \frac{|\alpha|}{2} \times \\
    &\mathcal{W}\left(\gamma+\frac{\alpha}{2},\gamma_b-\alpha \frac{k'(\gamma)}{k_b'(\gamma_b)};\gamma-\frac{\alpha}{2},\gamma_b+\alpha \frac{k'(\gamma)}{k_b'(\gamma_b)}\right)
\end{aligned}
\end{equation}
and find that the collision integral indeed takes the form~\eqref{eq:Igrazing} upon truncating the Dirac delta expansion after the second term.

\section{Derivation of collision operator from GBBGKY hierarchy}\label{app:Q-derivation}
The following section is based heavily on formalism and notation introduced in~\cite{biagetti2024generalised}, especially on the part related to many-tube systems.
Thus, for the purpose of clearer presentation in this section we change the notation slightly. The quasimomentum dependence will be denoted in lower index and we introduce rotation matrices $R=\mathbf{1}- \mathcal{T}n$. We also denote $\mathcal{A}_{\lambda_1,\lambda_2} = R^{-t}_{\lambda_1,\gamma}v_\gamma R^{t}_{\gamma,\lambda_2}$ and $\mathcal{C}_{\lambda_1,\lambda_2}=R^{-t}_{\lambda_1,\gamma}\rho_\gamma f_\gamma R^{-1}_{\gamma,\lambda_2}$. The upper indices $(i)$ denote tube index. To system tube we will refer to with index $(1)$, bath tube will be denoted with $(2)$. Einstein summation convention will be assumed in integration over repeated quasimomenta.
We start with the following set of equations:
\begin{equation}
\begin{split}
    \partial_t \rho^{(1)}_{\lambda_1} &= \int \dd x_2 \dd \lambda_2 V'(x_1-x_2)\partial_{\lambda_1} g_{\lambda_1,\lambda_2}(x_1,x_2)\\
    \partial_t \rho^{(2)}_{\lambda_2} &= \int \dd x_1 \dd \lambda_1 V'(x_1-x_2)\partial_{\lambda_2} g_{\lambda_1,\lambda_2}(x_1,x_2)
\end{split}
\end{equation}
\begin{equation}
\begin{aligned}
    &\partial_t g_{\lambda_1, \lambda_2}(x_1,x_2) +\\
    &\mathcal{A}_{\lambda_1,\gamma}^{(1)}\partial_{x_1}g_{\gamma,\lambda_2}(x_1,x_2)+\mathcal{A}_{\lambda_2,\gamma}^{(2) }\partial_{x_2}g_{\lambda_1,\gamma}(x_1,x_2)=\\
    &V'(x_1-x_2) \int \dd \lambda_3 \left(\partial_{\lambda_1}\rho_{\lambda_1}^{(1)}\mathcal{C}^{(2)}_{\lambda_2,\lambda_3}-\partial_{\lambda_2}\rho_{\lambda_2}^{(2)}\mathcal{C}^{(1)}_{\lambda_1,\lambda_3}\right)
\end{aligned}
\end{equation}
We consider homogenous system with $z=x_1-x_2$ 
\begin{equation}\label{eq:1pointfun}
\begin{split}
    \partial_t \rho^{(1)}_{\lambda_1} &= \int \dd z \dd \lambda_2 V'(z)\partial_{\lambda_1} g_{\lambda_1,\lambda_2}(z)\\
    \partial_t \rho^{(2)}_{\lambda_2} &= \int \dd z \dd \lambda_1 V'(z)\partial_{\lambda_2} g_{\lambda_1,\lambda_2}(z)
\end{split}
\end{equation}
\begin{equation}\label{eq:2pointfun}
    \partial_t g_{\lambda_1,\lambda_2} + \partial_z \left( \mathcal{A}^{(1)}_{\lambda_1,\gamma}g_{\gamma,\lambda_2}-\mathcal{A}_{\lambda_2,\gamma}^{(2)}g_{\lambda_1, \gamma}\right)=\mathcal{S}_{\lambda_1,\lambda_2}
\end{equation}
with source term 
\begin{equation}
    \mathcal{S}_{\lambda_1,\lambda_2}=V'(z)\int \dd \lambda_3 \left(\partial_{\lambda_1} \rho_{\lambda_1}^{(1)}\mathcal{C}^{(2)}_{\lambda_2, \lambda_3} - \partial_{\lambda_2}\rho_{\lambda_2}^{(2)}\mathcal{C}^{(1)}_{\lambda_1,\lambda_3}\right)
\end{equation}
To make progress with \eqref{eq:2pointfun} we rotate to normal modes of GHD as
\begin{equation}
  g_{\lambda_1,\lambda_2}=(R^{(1)})^{-t}_{\lambda_1,\gamma_1} (R^{(2)})^{-t}_{\lambda_2,\gamma_2} \bar{g}_{\gamma_1,\gamma_2},  
\end{equation}
which leads to (below we define $v_{\lambda_1,\lambda_2}:=v^{(1)}_{\lambda_1}-v^{(2)}_{\lambda_2}$)
\begin{equation}
    \partial_t \bar{g}_{\lambda_1,\lambda_2} + v_{\lambda_1,\lambda_2}\partial_z \bar{g}_{\lambda_1,\lambda_2}= \bar{S}_{\lambda_1,\lambda_2}.
\end{equation}
This equation is a linear equation with a source term, thus it can be solved easily. The Green's function is
\begin{equation}
    \mathcal{G}_{\lambda_1,\lambda_2}(z,z';t,t') = \frac{1}{2 \pi} \int \dd k \, e^{ik(z+z'-v_{\lambda_1,\lambda_2}t)}\theta_H(t-t')
\end{equation}
with $\theta_H$ denoting Heaviside function. The solution is given by
\begin{equation}
    \bar{g}_{\lambda_1,\lambda_2}(z,t) = \int \dd z' \dd t' \mathcal{G}_{\lambda_1,\lambda_2}(z,z';t,t') \bar{\mathcal{S}}_{\lambda_1,\lambda_2}(z',t').
\end{equation}
After rotating back we get
\begin{equation}\label{eq:g21solution}
\begin{aligned}
    &g_{\lambda_1, \lambda_2}[\rho](z,t) =(R^{(1)})_{\lambda_1,\gamma_1}^{-t} (R^{(2)})_{\lambda_2,\gamma_2}^{-t}  \int_0^t \dd \tau  \int \frac{\dd k}{2 \pi} \int \dd z' \times \\
    &e^{ik(z+z'-v_{\gamma_1, \gamma_2}\tau)}(R^{(1)})_{\gamma_1,\eta_1}^t (R^{(2)})_{\gamma_2,\eta_2}^t \mathcal{S}_{\eta_1, \eta_2}[\rho](z',t-\tau).
\end{aligned}
\end{equation}
In the next step we plug this expression into $\eqref{eq:1pointfun}$, and perform the standard approximations used in such derivations~\cite{Balescu1997}, related to separations of scales. In practice replace the time argument of $\mathcal{S}_{\eta_1, \eta_2}[\rho](z',t-\tau)$ with $t$ and extend the integration range over $\tau$ to infinity. We also insert momentum representations of interaction potential $V(z) = \int \dd k e^{ikz} \tilde{V}(k)$ and use the property that $\int \dd \lambda R^{-t}_{\lambda,\gamma}=2 \pi \rho^{\rm t}_\gamma$. After all these steps we find the collision integral
\begin{equation}
	\partial_t \rho_{\lambda_1}^{(1)} = \mathcal{I}(\lambda_1), \qquad \mathcal{I}(\lambda_1) = \partial_{\lambda_1} \left[(R^{(1)})^{-t}_{\lambda_1,\eta} \mathcal{I}_0(\eta)\right],
\end{equation}
where 
\begin{widetext}
\begin{equation}\label{eq:BoltzmannGGBKY}
	\mathcal{I}_0(\lambda_1) = (2 \pi)^2 \int \dd k k^2 \tilde{V}^2(k)  \int_0^\infty \dd \tau\int\dd \lambda_2 e^{-ik v_{\lambda_1 \lambda_2 }\tau}\rho^{\rm t,(2)}_{\lambda_2}(R^{(1)})_{\lambda_1,\eta_1}^t (R^{(2)})_{\lambda_2,\eta_2}^t \int \dd \eta_3 \left(\partial_{\eta_1} \rho_{\eta_1}^{(1)} \mathcal{C}^{(2)}_{\eta_2 \eta_3} - \partial_{\eta_2}\rho_{\eta_2}^{(2)}\mathcal{C}^{(1)}_{\eta_1\eta_3}\right).
\end{equation}

Where by $\rho^{\rm t,(2)}_{\lambda_2}$ we denoted total density of states in tube 2. We proceed by using~\cite{Balescu1997} $\int_0^{\infty} \dd \tau e^{i \tau x} = \pi \delta(x)+i \mathcal{P}(\frac{1}{x})$, where the second term is the principal value, which vanishes as the integrand is a symmetric function in $k$. In what follows we also use that~\cite{biagetti2024generalised} $\int \dd \lambda_2\, \C_{\lambda_1,\lambda_2}=2 \pi R_{\lambda_1,\gamma}^{-t} \rho_\gamma f_\gamma \rho^{\rm t}_\gamma$ and $ R^t_{\lambda,\gamma}\partial_\gamma \rho_\gamma =- \rho_\lambda f_\lambda \partial_\lambda \epsilon_\lambda$. We get
\begin{equation}
    \mathcal{I}_0(\lambda_1) = - 2 \pi^2 \int \dd k |k|\tilde{V}^2(k) \int \dd \lambda_2 \delta(v_{\lambda_1,\lambda_2})  \left(k_2'(\lambda_2)\right)^2 \rho_{\lambda_1}^{(1)} f_{\lambda_1}^{(1)} \rho_{\lambda_2}^{(2)} f_{\lambda_2}^{(2)} \left(\epsilon'_{\lambda_1}- \frac{k'_1(\lambda_1)}{k'_2(\lambda_2)} \epsilon'_{\lambda_2}\right).
\end{equation}
In the end we find collision integral which has essentially  exactly the same stationary state~\eqref{eq:stat_grazing} as the one found in the grazing limit of collision integral~\eqref{eq:Master_general}. 

\end{widetext}

\section{Molecular dynamics simulation algorithm}\label{app:algorithm}

The classical system with collision integral given by~\eqref{eq:swap1} can be simulated fairly easy with a standard implementation of Gillespie algorithm~\cite{Gillespie1977}. The fermionic case~\eqref{eq:swap2} is more complicated due to Fermi exclusion principle, which we want to somehow model using dynamics of classical particles. We thus wish to construct a classical model for which the dynamics in phase-space would be described by kinetic equation~\eqref{eq:swap2}. Our approach will turn out to be more general and cover both classical and fermionic case.

To account for spatial locality of the collision integral~\eqref{FermiClassicalIntEqn} we begin by dividing the system into sufficiently small cells of equal length and consider each cell as a separate system, while allowing the particles to flow between different cells.
In the molecular dynamics simulation for the velocity-swap models we only scatter system particles with heat bath particles.
For the classical case, we have no restriction on the momenta of the scattered particles, so number of such collisions in each cell per unit time (rate) is
\begin{equation}
    R = \frac{\gamma \cdot N \cdot N_{\rm bath}}{L / ({\rm Number\;of\;cells})},
\end{equation}
where $N_{\rm}$ and $N_{\rm bath}$ are respectively number of system particles and number of bath particles in the cell and $L$ is total length of the system.
In the fermionic case we need to account for the Pauli exclusion principle, which prohibits scattering if we already have a particle with the same (outgoing) momentum.
Consequently, scattering rate needs to be modified for fermions:
\begin{equation}
    R = \frac{\gamma \cdot N \cdot |\{v_{b,i} \notin V_{\rm system} : i \in \{1, ..., N_{\rm bath}\}\}|}{L / ({\rm Number\;of\;cells})},
\end{equation}
where by $V_{\rm system}$ we understand set of velocities of system particles in the cell and $v_{b,i}$ denotes velocity of $i$-th bath particle in the cell.
For fermions we need to be careful, as dividing the system into cells we ran into a situation where the ballistic propagation of a particle can cause the exclusion principle to be violated.
For the simulation presented in this paper, we resolve this problem by artificially scattering particles violating the exclusion principle with the particles from the bath. For the parameters used for the simulation presented in Fig.~\ref{MolecularDynamicsPlots}, the described situation is rare and we observed that no matter how this issue is resolved, density profiles do not change significantly.

In the simulations shown in Fig.~\ref{MolecularDynamicsPlots} we performed the following variation of discrete step Monte Carlo algorithm~\cite{Bird1994} with a fixed time step, which we denote $\Delta t$:
\begin{enumerate}
    \item Segregate all system particles into cells of equal size.
    \item For each cell:
    \begin{enumerate}
        \item Compute rate $R$.
        \item If we find exclusion principle to be violated (due to particle streaming), we scatter duplicates with the particles from the bath, so that there are no two particles with the same velocity in each cell.
        \item Draw a random number from 0 to 1. If $R \cdot \Delta t$ is greater then the drawn number, go to next cell.
        \item Otherwise, randomly choose system particle and bath particle in the cell and set the momentum of the system particle to the momentum of the bath particle.
    \end{enumerate}
    \item Ballistically propagate system particles by a time $\Delta t$ and go to the step 1.
\end{enumerate}

\section{Additional formulas and derivations} \label{app:additional}

In this Appendix we present additional computations showing: Markovianity of the GHD diffusion operator, existence and uniqueness of the solution to the ChE equation and verify that parameter $u$ transforms as a velocity while the diffusion coefficient is invariant under the Galilean transformations.

\subsection{Diffusion operator and its Markovianity}\label{app:diffusion}

We recall here the result of showing the GHD diffusion operator to be Markovian~\cite{DeNardis2019}, namely
\begin{equation} \label{app:Markov_diff}
    \int {\rm d}\lambda\, \mathfrak{D}_\rho(\lambda, \lambda') = 0.
\end{equation}
The GHD diffusion operator is~\cite{DeNardis2018,Gopalakrishnan2018} 
\begin{align} \label{eq:ghd_diff_operator_form}
    \mathfrak{D}_{\rho} &=
       \left(\mathbf{1} - n \mathcal{T}\right)^{-1} \rhot^{-1}
            \Tilde{\mathfrak{D}}_{\rho}
            \rhot^{-1} \left(\mathbf{1} - n \mathcal{T}\right), \\
   \Tilde{\mathfrak{D}}_{\rho}(\lambda, \lambda') &=
        \delta\left(\lambda - \lambda'\right) w(\lambda) - W(\lambda, \lambda'),
\end{align}
with
\begin{align}
    W(\lambda, \lambda') = \rhop(\lambda)f(\lambda)
        \left(\mathcal{T}^{\mathrm{dr}}(\lambda, \lambda')\right)^2
        \left|v_{\rho}(\lambda) - v_{\rho}(\lambda')\right|,
\end{align}
and $w(\lambda) = \int \dd \lambda' W(\lambda', \lambda)$. We note that the operator $(\mathbf{1} - n\mathcal{T})^{-1}$ implements dressing ${\rm dr}$ when acting to the left,
\begin{equation}
    f^{\rm dr} = f \cdot (\mathbf{1} - n \mathcal{T})^{-1}.
\end{equation}

Relation~\eqref{app:Markov_diff} follows from the fact that $2\pi \rho_{\rm tot} = (h_0)^{\rm dr}$. This gives
\begin{equation}
    \frac{1}{2\pi}\int {\rm d}\lambda \mathfrak{D}_\rho(\lambda, \lambda') = \int {\rm d}\lambda \left(\tilde{\mathfrak{D}}_\rho \rho_{\rm t}^{-1}(\mathbf{1} - n \mathcal{T})\right)(\lambda, \lambda'),
\end{equation}
and the latter integral vanishes because of the structure of $\Tilde{\mathfrak{D}}_{\rho}(\lambda, \lambda')$.

\subsection{Existence and uniqueness of the diffusion coefficient} \label{app:existence}

The diffusion coefficient is expressed by the solution to the integral equation~\eqref{int_eq_ChE}. We show here that the solution to this equation exists and is unique. This follows from the standard result in the theory of linear integral equations which we now recall following~\cite{Resibois}.

The equation in question is a linear inhomogeneous integral equation of the form
\begin{equation}
    L \cdot a = b.
\end{equation}
Let us assume that $L$ is hermitian. The solution to such equation exists if $b$ is orthogonal to all solutions $\phi_j$ of the homogeneous equation
\begin{equation}
    L \cdot \phi_j = 0.
\end{equation}

In the present context, there is only one solution, $\phi = {\rm const}$, to the homogeneous equation $\Gamma^{\rm st} \cdot \phi = 0$. It corresponds to the single conserved quantity, that is the particle number. Therefore, we need to check that
\begin{equation}
    \mathbf{1} \cdot \mathcal{A}^{\rm st} \cdot \frac{\partial \rhop^{\rm st}}{\partial \varrho} = u,
\end{equation}
where we used that
\begin{equation}
    \mathbf{1} \cdot \frac{\partial \rhop^{\rm st}}{\partial \varrho} = \frac{\partial \varrho}{\partial \varrho} = 1.
\end{equation}
From the definitions of the hydrodynamic operator $\mathcal{A}$ and the effective velocity we have
\begin{equation}
    \mathbf{1} \cdot \mathcal{A}^{\rm st} = (\ulb{1}')^{\rm dr} v_\rho \cdot (\mathbf{1} - n \mathcal{T}) = \ulb{2}' = \ulb{1}. 
\end{equation}
In the second equality we used that $v_\rho = (\ulb{2}')^{\rm dr}/(\ulb{1}')^{\rm dr}$ and that $(\mathbf{1} - n \mathcal{T})$ removes the dressing from the function that it acts upon. Recalling the definition~\eqref{eq:xi_def} of $u$ we confirm the sought after relation. This proves that if the operator $\Gamma^{\rm st}$ is hermitian then there is always a solution to the equation specifying the diffusion coefficient. The operators $\Gamma^{\rm st}$ obtained from the Fermi's golden rule or from RTA are hermitian which concludes the existence part of the proof.

The constitutive relation~\eqref{constitutive_relation} guarantees the uniqueness of the solution by removing from the space of solutions cases that differ by the kernel of $\Gamma^{\rm st}$.

\subsection{Galilean transformations}\label{app:galilean}

The Galilean transformations describe the effect of the boost, by a constant velocity $w$, on the system. We denote with {\em bar} quantities referring to the system moving with velocity $w$ with respect to the stationary one.

The quasiparticle distribution functions of the two systems obey $\bar{\rho}_{\rm p}(\lambda) = \rhop(\lambda - w)$. 
The parameter $u$ transforms as velocity under the boost,
\begin{equation}
    \bar{u} = u + w, 
\end{equation}
as can be directly verified from its definition~\eqref{eq:xi_def} and the transformation rule for the quasiparticles distributions. In a similar fashion we find transformation rules for the effective velocity and the flux Jacobian transform
\begin{equation}
    \begin{aligned}
        \bar{v}_{\rho}(\lambda) &= v_{\rho}(\lambda - w) + w,\\
        \bar{\mathcal{A}}(\lambda, \lambda') &= \mathcal{A}(\lambda - w, \lambda' - w) + w.
    \end{aligned}
\end{equation}
This implies that the solutions ChE equation~\eqref{int_eq_ChE} in the moving and in the stationary system are related by $\bar{a}(\lambda) = a(\lambda - w)$. The diffusion coefficient in the moving system is then
\begin{equation}
    \bar{\mathcal{D}} = - \int {\rm d}\lambda\, \ulb{1}(\lambda) \bar{a}(\lambda) = \mathcal{D} - \int {\rm d}\lambda\, a(\lambda).
\end{equation}
The remaining integral is zero due to the constitutive relation~\eqref{constitutive_relation}. This shows that the diffusion coefficient is invariant under the Galilean transformations.

\section{Linearization of the collision integral} \label{app:linearization}

In this section we approach computation of the linearized collision integral operator $\st{\Gamma}$~\eqref{LinearizedFullDef} for the two coupled Lieb--Liniger tubes.
Our derivation resembles linearization performed in~\cite{navierstokes}, although the context is a little different.
We assume the inter-tube interaction potential (in the momentum space) to have a form
\begin{equation}
    \tilde{V}(k) = \mathcal{V}(k/k_r), \quad \tilde{V}(k) = \tilde{V}(-k).
\end{equation}
Constant $k_r$ is a characteristic momentum exchanged between the tubes, which we assume to be small. It can also be regarded as an inverse of characteristic range of the potential.
In the small momentum transfer limit, processes involving 1 particle-hole excitation are dominant and higher processes contribute at higher powers of $k_r$~\cite{Lebek2024}.
While it is possible to directly compute $\delta \mathcal{I}[\rhop]/\delta \rhop$, such path is not the easiest way to find the linear contribution to the collision integral.
The task becomes much easier if instead of describing the perturbation with $\delta \rhop$ we switch the description to pseudoenergy perturbation.
In the language of pseudoenergy perturbation we define the linearized collision integral as 
\begin{equation}
    \st{\tilde{\Gamma}} =
        \st{\mathbb{F}_1} \cdot \frac{\delta \mathcal{I}_{0}}{\delta \epsilon_0}\Big|_{\st{\rhop}} .
\end{equation}
Using the identities~\cite{navierstokes}
\begin{align}
\begin{split} \label{EnergyDesityPerturbationRelations}
    \delta \rhop = -\C \cdot \delta \epsilon_0, \qquad
    \delta \epsilon = \left(\mathbf{1} - \mathcal{T} n\right)^{-1} \cdot \delta \epsilon_0
\end{split}
\end{align}
and remembering that $\mathcal{I}_0[\st{\rhop}] = 0$, we can relate $\st{\tilde{\Gamma}}$ to $\st{\Gamma}$ (see~\eqref{LinearizedFullDef}).
The presented equalities follow directly from the definition of the pseudoenergy and the equation~\eqref{eq:gge_condition}.
For arbitrary functions $f$ and $g$ we have
\begin{equation} \label{TildeGamma_not_expanded}
    \langle f|\st{\tilde{\Gamma}} g\rangle =
        \int \dd \lambda  \Drst{f}\, \frac{\delta \mathcal{I}_{0}}{\delta 
            \epsilon_0}\Big|_{\st{\rhop}}\cdot \st{g}.
\end{equation}
Writing the matrix element we used the identity
    $\int \dd \lambda f \cdot \mathbb{F} \cdot g = \int \dd \lambda f^{\mathrm{Dr}} \cdot g$, which follows from the definition of the operator $\mathbb{F}$~\cite{navierstokes}.
Given a state in the non-evolving tube and a particle number in the evolving tube, stationary state in the evolving tube is fixed with the condition~\eqref{stationarity}.
As discussed in the main text, to linearize collision integral in the density imbalance limit we consider a perturbation $\rhop = \st{\rhop} +\delta \rhop$, while keeping the state in the non-evolving tube $\rhop^{(b)}$ fixed.
Combining~\eqref{TransitionRates} with the dynamic structure  factor~\eqref{DynamicStructureFactor} we write $\mathcal{I}_0$ in a form convenient for linearization
\begin{align*}
    \mathcal{I}_{0}&[\rhop](p_1) = (2\pi)^2 \int
        \dd h_1 \,
        \dd p_2 \, \dd h_2 \\
        & \times G(p_1, h_1, p_2, h_2) \,
        H(p_1, h_1, p_2, h_2),
\end{align*}
where
\begin{align*}
    G = &\tilde{V}^2(k(p_1, h_1)) \rhop(h_1) \, \rhoh(p_1) \, \rhop^{(b)}(p_2) \rhoh^{(b)}(h_2)\\
        & \times |F(p_1, h_1)|^2 |F^{(b)}(p_2, h_2)|^2 \\
        & \times \delta(k(p_1, h_1) + k^{(b)}(p_2, h_2)) \\
        & \times \delta(\omega(p_1, h_1) + \omega^{(b)}(p_2, h_2)), \\
    H &=
        \exp\left(\epsilon(p_1) - \epsilon(h_1)
            + \epsilon^{(b)}(p_2) - \epsilon^{(b)}(h_2)\right) - 1.
\end{align*}
We perturb $\rhop$ around stationary state and linearize. It follows that
\begin{align*}
    \langle f|\st{\tilde{\Gamma}} g\rangle &= (2 \pi)^2 \int \dd p_1 \, \dd h_1 \, \dd p_2 \, \dd h_2 \, \Drst{f}(p_1)  \\
        & \times \st{G}(p_1, h_1, p_2, h_2) \left(\Drst{g}(p_1) - \Drst{g}(h_1)\right),
\end{align*}
where we performed the trivial integral in $\lambda$.
Next step is to change coordinates to the center of mass (CM) coordinates
\begin{equation}
    \lambda_i = \frac{p_i + h_i}{2} \quad \mathrm{and} \quad \mu_i = p_i - h_i,
\end{equation}
with the Jacobian of change of variables being unity.
Working in the small momentum limit, $\mu_i$ is small.
In the CM variables Dressed momentum and energy are
\begin{align*}
    k(p, h) &= k'(\lambda) \mu + \bigO\left(\mu^3\right), \\
    \omega(p, h) &= \omega'(\lambda) \mu + \bigO\left(\mu^3\right).
\end{align*}
Also, in small momentum limit form factors simplify to
    $F(p, h) = k'(\lambda) + \bigO(\mu^2)$~\cite{Panfil2023}.
The interaction potential term is expanded as
\begin{align}
    \tilde{V}^2(k(p) - k(h)) &=
        \tilde{V}^2\left(k'(\lambda) \mu\right) + \bigO\left(\mu^3\right).
\end{align}
Expanding $\st{G}$ and keeping terms up to order $\bigO(\mu^2)$
\begin{align*}
    \st{G} &= \tilde{V}^2\left((\st{k})'(\lambda_1) \mu_1\right)
       (\st{k})'(\lambda_1) \, (k^{(b)})'(\lambda_2) \\
    & \times \st{\rhop}(\lambda_1)\st{\rhoh}(\lambda_1)\rhop^{(b)}(\lambda_2)\rhoh^{(b)}(\lambda_2) \\
    & \times \frac{1}{|\mu_1|}
        \delta\left(\mu_2 + \mu_1 \frac{(\st{k})'(\lambda_2)}{(k^{(b)})'(\lambda_1)}\right) \\
    & \times \delta\left(\st{v}(\lambda_1) - v^{(b)}(\lambda_2)\right),
\end{align*}
where $v = \omega'/k'$ is the effective velocity~\eqref{eq:eff_v_int_eq}.
Changing variables under the Dirac delta we used that $k' = 2 \pi \rhot > 0$.
Plugging expanded $\st{G}$ into~\eqref{TildeGamma_not_expanded} we find
\begin{align*}
    \langle f|\st{&\tilde{\Gamma}} g\rangle = (2 \pi)^2 \int \dd \lambda_1 \, \dd \mu_1 \, \dd \lambda_2 \,
      \tilde{V}^2\left((\st{k})'(\lambda_1) \mu_1\right) \\
        & \times (\st{k})'(\lambda_1) \, (k^{(b)})'(\lambda_2)
            \st{\rhop}(\lambda_1)\st{\rhoh}(\lambda_1)\rhop^{(b)}(\lambda_2)\rhoh^{(b)}(\lambda_2) \\
        & \times \frac{1}{|\mu_1|}
        \delta\left(\st{v}(\lambda_1) - v^{(b)}(\lambda_2)\right)
         \\
        & \times \mu_1 \, \left(\Drst{g}(\lambda_1)\right)' \left(\Drst{f}(\lambda_1) + \mu_1 \drst{\left(f'\right)}(\lambda_1)\right).
\end{align*}
Writing the matrix element we took advantage of the relation between dressings~\eqref{dressing_relation}.
Proceeding we note that since $\tilde{A}$ is symmetric in momenta, the term
    $\sim \Drst{f}$ does not contribute to the matrix element.
Changing variables $\mu_1 \to \mu_1 / k'$, after some bookkeeping we arrive at
\begin{align*}
    \langle f|\st{&\tilde{\Gamma}} g\rangle = (2 \pi)^2  \, k_r^ 2 \, |\mathcal{V}^2| \\
    & \times \int \dd \lambda \,
        \frac{\st{\rhop} \st{\rhoh}}{(\st{k})'} \, \zeta^{(b)}\left(\st{v}\right) 
      \drst{\left(f'\right)} \left(\Drst{g}\right)' + \bigO(k_r^3),
\end{align*}
where $ |\mathcal{V}^2| = \int \dd z \, |z| \, \mathcal{V} ^2(z)$ and
\begin{equation}
    \zeta^{(b)}(v) = \frac{\rhop^{(b)} \, \rhoh^{(b)} \, (k^{(b)})'}{|(v^{(b)})'|}\Bigg|_{\left(v^{(b)}\right)^{-1}\left(v\right)}.
\end{equation}
Finally we note that in the Tonks--Girardeau limit ($c \to \infty$) in both
    tubes, the matrix element takes particularly simple form
\begin{equation}
    \langle f|\st{\tilde{\Gamma}} g\rangle
        = (2 \pi)^2  \, k_r^ 2 \, |\mathcal{V}^2| \int \dd \lambda \,
        \st{\rhop} \st{\rhoh} \rhop^{(b)} \rhoh^{(b)} \,
        f' \, g',
\end{equation}
since in this limit the dressing is absent.

\section{Additional numerical results and verifications} \label{app:numerical}

In this section we describe numerical computations of the diffusion
    coefficient for the two density-imbalanced coupled Lieb--Liniger tubes.
First, note that the only non-trivial step needed to compute diffusion coefficient is
    solving the integral equation~\eqref{int_eq_ChE}.
The equation is an example of a Fredholm integral equation of the first kind.
A simple way to solve it is by choosing a basis and restricting it to
    a finite dimension.
Using~\eqref{EnergyDesityPerturbationRelations} we rewrite~\eqref{int_eq_ChE} into equivalent form
\begin{equation}
    \st{\tilde{\Gamma}} \cdot \alpha = \left(\st{\mathcal{B}} - u \st{\mathcal{C}}\right)
        \cdot \partial_{\varrho} \epsilon_0,
\end{equation}
where $\Cst \cdot \alpha = a$.
Continuing, we write $\alpha$ in a polynomial basis, orthonormal with respect to hydrodynamic scalar product
\begin{equation} \label{eq:hydrodynamic_scalar_product}
    \hst{f}{g} = \int \dd \lambda \, f(\lambda) \left(\Cst \cdot g\right)(\lambda).
\end{equation}
The basis is constructed by performing Gram-Schmidt orthonormalization on the ultra-local charge
    basis~\cite{navierstokes}.
We will denote elements of the new basis as $\{\mathfrak{h}_i\}_{i \in \mathbb{N}}$.
Note, that from the construction procedure follows $\mathfrak{h}_0 \sim \ulb{0}$.
Expanding $\alpha = \alpha_j \, \mathfrak{h}_j$, multiplying from the left by $h_i$ and integrating, we arrive at
\begin{equation}
    \st{\tilde{\Gamma}}_{ij} \, \alpha_j = \st{B}_i - u \st{C}_i,
\end{equation}
where Einstein's summation convention is assumed.
Matrix elements are defined as
\begin{align}
\begin{split} \label{Gamma_matrix_element}
    \st{\tilde{\Gamma}}_{ij} &= \langle \mathfrak{h}_i|\tilde{\st{\Gamma}} \mathfrak{h}_j\rangle, \\
    \st{B}_i &= \int \dd\lambda \, v_{\st{\rhop}} \, \st{\rhop} \, \st{f} \,  \drst{\mathfrak{h}_i} \,
        \drst{\left(\partial_{\varrho} \epsilon_0\right)}, \\
    \st{C}_i &= \int \dd\lambda \, \st{\rhop} \, \st{f} \,  \drst{\mathfrak{h}_i} \,
        \drst{\left(\partial_{\varrho} \epsilon_0\right)},
\end{split}
\end{align}
Straightforward way to solve this set of equations is to truncate the basis to a finite dimension.
Inverting the matrix $\st{\tilde{\Gamma}}_{ij}$ we encounter a complication.
Because of the particle
    conservation law~\eqref{eq:boltzmann_conservation} we have $\st{\tilde{\Gamma}}_{0,j} = 0$.
Thus the matrix is rank deficient.
From computations performed in the section~\ref{app:existence} follows $\st{B}_0 - u \,\st{C}_0 = 0$.
To be able to uniquely determine $\alpha_i$, one needs to use the constitutive relation~\eqref{constitutive_relation}, which fixes $\alpha_0 = 0$, resolving the rank deficiency problem.
Higher components of $\alpha$ are fixed by the matrix equation
\begin{equation} \label{MatrixEquationForA}
    \mathbb{\st{\tilde{\Gamma}}} \cdot \vec{\alpha} = \st{\vec{B}} - u \, \st{\vec{C}},
\end{equation}
where $(\mathbb{\st{\tilde{\Gamma}}})_{ij} = \st{\tilde{\Gamma}}_{i + 1,j + 1}$ is a full-rank matrix and $(\st{\vec{B}})_{i} = \st{B}_{i + 1}$, $(\st{\vec{C}})_{i} = \st{C}_{i + 1}$, $(\vec{\alpha})_i = \alpha_{i + 1}$.
We have verified that solution to~\eqref{MatrixEquationForA}, upon increasing the basis size, converges to function solving the original integral equation. 

We applied the presented method to the system of weakly coupled Lieb--Liniger tubes.
Data presented in Fig.~\ref{TwoTubesPlot} were computed using the presented method.

We now turn to analyzing the function $\partial_\varrho \epsilon_0$ for the system of two Lieb--Liniger tubes.
Suppose the environmental tube is in a boosted thermal state.
Pseudo energy is then $\epsilon^{(b)}(\lambda) = -\beta \, \mu^{(b)} + \kappa \, k^{(b)}(\lambda) + \beta \, \omega^{(b)}(\lambda)$, where $\beta = 1/T$ and $\kappa$ is a potential related with the tube's velocity. 
In~\cite{Lebek2024} it is shown that for the environmental tube in a boosted thermal state, the pseudoenergy of the stationary state in the coupled tube is $\st{\epsilon}(\lambda) = -\beta \, \mu + \kappa \, \st{k}(\lambda) + \beta \, \st{\omega}(\lambda)$.
Consequently, the bare pseudo energy reads $\st{\epsilon_{0}}(\lambda) = \beta \, \mu + \kappa \, \ulb{1}(\lambda) + \beta \, \ulb{2}(\lambda)$.
To compute the bare pseudoenergy we differentiated~\eqref{eq:gge_condition} and used the relation between
    dressing operations~\eqref{dressing_relation}.
We see that the only free parameter is the chemical potential $\mu$, fixing number of particles
    in the evolving tube.
Therefore, applying the chain rule,
    $\partial_\varrho \st{\epsilon_{0}} = \frac{\dd \mu}{\dd \varrho}
        \partial_{\mu} \st{\epsilon_{0}} = -\frac{\dd \mu}{\dd \varrho} \beta$.
The derivative $\frac{\dd \mu}{\dd \varrho}$ can be further simplified
\begin{equation}
    \frac{\dd \varrho}{\dd \mu} = \int \dd \lambda \frac{\partial \st{\rhop}}{\partial \mu}
        = -\int \dd \lambda \, \st{\C} \cdot \frac{\partial \st{\epsilon_{0}}}{\partial \mu}
        = -\beta \,\hst{\ulb{0}}{\ulb{0}},
\end{equation}
where in the second equality we used~\eqref{EnergyDesityPerturbationRelations}.
Last equality follows from the definition of the hydrodynamic scalar product.
Thus, for a system near a boosted thermal state
\begin{equation}
    \partial_\varrho \st{\epsilon_0} = \frac{1}{\hst{\ulb{0}}{\ulb{0}}}.
\end{equation}
For an athermal stationary state in the bath tube and finite interaction strength, the factor
    $\partial_\varrho \st{\epsilon_0}$ has much more complicated structure.
To compute it we begin by writing $\st{\epsilon}(\lambda)$ as $(\st{\epsilon})'(\lambda) + \nu$, where $\nu$ is a constant parameter.
For a fixed value of $\nu$, using~\eqref{stationarity} we are able to find $\st{\epsilon}(\lambda)$ by solving the non-linear integral equation.
Having solved equations for sufficiently many values of $\nu$ we are able to find $\varrho(\nu)$ and compute $\partial_{\varrho}\st{\epsilon_0}(\lambda)$.

\begin{figure}[t]
    \includegraphics[scale=0.5]{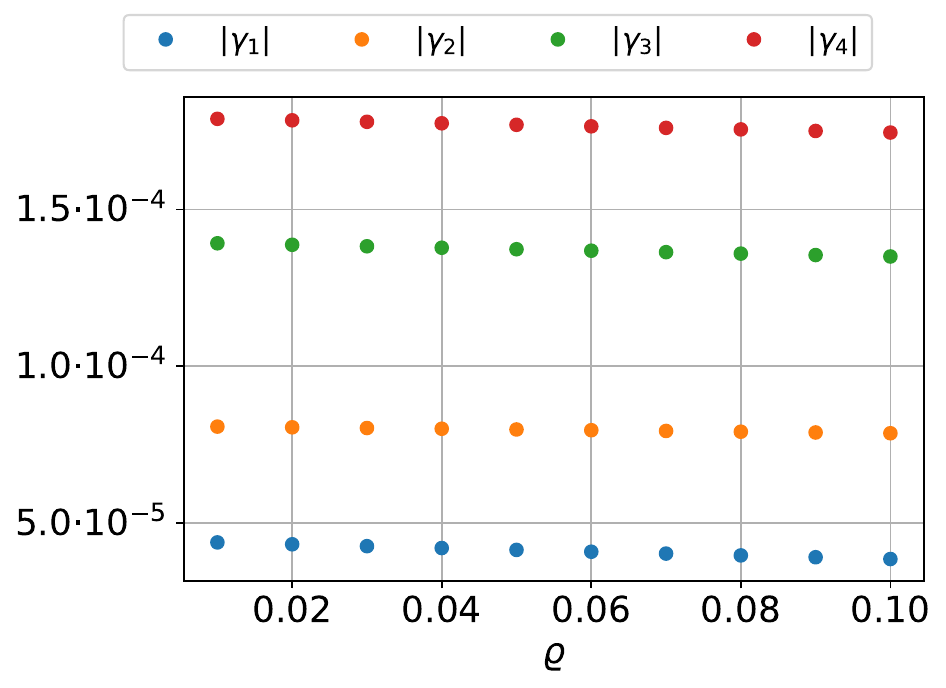}
    \caption{The smallest non-zero eigenvalues of $\st{\tilde{\Gamma}}_{ij}$ as a function of $\varrho$ for the system in the same state as in panel (a) in Fig.~\ref{TwoTubesPlot}.}
    \label{fig:gap}
\end{figure}

Finally, let us comment on the estimation of the relaxation time.
As discussed in the main text it is customary to estimate relaxation time of $\st{\Gamma}$ as a gap between zero and the smallest non-zero eigenvalue, see Fig.\ref{fig:gap}.
To numerically find the smallest non-zero eigenvalue of the operator $\st{\Gamma}$, algorithm similar to the one outlined above might be used.
It is straightforward to verify that if $\gamma$ is an eigenvalue of $\st{\Gamma}$, then it satisfies the equation
\begin{equation}
    \st{\tilde{\Gamma}}_{ij} \, \alpha_j = -\gamma \, \alpha_i
\end{equation}
for some vector $\alpha_i$.
To find the gap, one can for example perform the power iteration procedure on the $\tilde{\mathbb{\Gamma}}$ matrix.
We employed this method to find the approximation of the relaxation time presented in Fig.~\ref{TwoTubesPlot}.
The presented method has its weaknesses. 
Firstly, while we are able to verify that that the found eigenvalue is indeed an eigenvalue of the operator $\st{\Gamma}$, by no means it allows us to verify that it indeed is the smallest eigenvalue of the integral operator.
Lastly, we observed that the method fails to give reasonable approximation for larger temperatures.
Detailed analysis showed that for some cases the eigenvector corresponding to the smallest eigenvalue is converging towards some function for small truncation dimensions.
As we further increase the dimension, we observe that this convergence is lost before we find satisfactory approximation of the eigenvector, an effect similar to parameter overfitting.

\bibliography{refs.bib}

\end{document}